# Strong lead-free bioinspired piezoceramics for durable energy transducers

Ruxue Yang[1], Temesgen Tadeyos Zate[2], Peiren Wang[1], Soumyajit Mojumder[1], Elo Overgaard Mogensen[2], Oriol Gavalda-Diaz[1], Zihe Li[3], Ajeet Kumar[3], James Roscow[3], Hamideh Khanbareh[3], Astri Bjørnetun Haugen[2], Florian Bouville[1]

[1] Centre for Advanced Structural Ceramics, Imperial College London, London, United Kingdom

[2] Department of Energy Conversion and Storage, Technical University of Denmark, Agnes Nielsens vej, Building 301, 2800 Kgs Lyngby, Denmark

[3] Department of Mechanical Engineering, University of Bath, Bath, United Kingdom.

## Abstract

Durable, high-performance and environmentally friendly lead-free piezoceramics are essential for next-generation sustainable energy transducers and electromechanical systems. However, piezoceramics are mechanically weak, which negatively impacts their working conditions and durability. A solution to improve their mechanical durability without sacrificing piezoelectric performance remains elusive. Here, we design a bioinspired brick-and-mortar $Bi_{0.5}Na_{0.5}TiO_3$ (BNT) ceramics with BNT bricks and silica mortar. This deliberate microstructure design yields a 2- to 3-fold increase in flexural strength and a 1.6- to 2-fold increase in fracture toughness compared with conventional BNT without sacrificing piezoelectric performance. These improvements translate into a 10 to 15-fold increase in ferroelectric fatigue lifetime and a 46% higher voltage output and slower degradation as energy transducers. These gains originate from anisotropic residual stress fields, revealed by Raman spectroscopy and XRD, which delay crack initiation events. Given its microstructural origin, this strategy is broadly applicable to other systems where both functional and structural performance are critical.

## Introduction

Polycrystalline piezoceramics are critical functional materials capable of directly converting electrical and mechanical energy.[1] They serve as core components in a wide range of electronic systems, including transducers (sensors, actuators), energy storage devices and energy harvesters.[1–3] This makes piezoceramics essential for aerospace, medical engineering, precision manufacturing, and intelligent systems applications.[1–4] The most dominant and performant piezoceramics products are lead-based materials, but their lead content raises environmental concerns during production and disposal.[1] This has driven intense efforts toward developing high-performance, lead-free alternatives, such as $Bi_{0.5}Na_{0.5}TiO_3$ (BNT), $BaTiO_3$ (BT) and $K_{0.5}Na_{0.5}NbO_3$ (KNN)-based ceramics. This drive to improve the overall piezoelectric, ferroelectric and electro-strictive strain performance of piezoceramics has been mainly focused on changing the composition, with innovative concepts such as morphotropic phase boundary (MPB)[5–7], domain wall engineering[1,8], domain engineering[9–11], local structural heterogeneity engineering[12,13], and introducing crystallographic texture by using single crystals or controlling grain orientation through tape casting and templated grain growth[14,15]. These concepts lead to impressive piezoelectric properties, using crystallographic texture in lead zirconate titanate (PZT) for instance, with piezoelectric charge coefficients ($d_{33}$) of up to 760 pC/N and high Curie temperature ($T_c$ ≈ 360 °C) in <100> textured PZT ceramics made with Templated Grain Growth[16]. This approach has been translated to lead-free systems, with excellent piezoelectric performance achieved in the lead-free KNN-based system ($d_{33}$ ≈ 700 pC/N, $d_{33}^*$ ≈ 980 pm/V, $k_p$ = 76%) enabled by ⟨001⟩ texturing through domain engineering[17]. Finally, the $Bi_{0.5}Na_{0.5}TiO_3$-$Sr_{0.7}Bi_{0.3}TiO_3$-based system with ⟨111⟩ texturing shows high dielectric breakdown strength (103 MV/m), enhanced fatigue resistance, and ultrahigh energy density (21.5 J/cm$^3$).[18] The link between piezo and mechanical properties goes even deeper, as ferroelastic materials are not linear elastic, with depolarisation under stress potentially introducing toughening mechanisms and non-linearity during mechanical testing.

A major limitation of both lead-based and lead-free piezoceramics is their intrinsic low mechanical properties, including both strength and fracture toughness[4,19]. This not only compromises material reliability by constraining operational load thresholds and increasing the risk of mechanical failure but also exacerbates electromechanical fatigue and dielectric breakdown from microcrack propagation under cyclic loading and high electric fields. This issue is particularly pronounced in multilayer actuators[4] and mechanical energy harvesters[20]. More than being an issue, however, studying and influencing this link between microstructure, functional and mechanical properties represents an important scientific question that could open new avenues to improve the performance and long-term durability of functional materials.

Previous strategies to increase the mechanical properties of piezoceramics have been based on conventional ceramic toughening approaches[4], such as adding second-phase particles[21–23], whiskers[24], or by tailoring grain size[25]. However, the mechanical enhancements have typically been made at the expense of piezoelectric performance[4], and studies on reliability and cyclic stability remain scarce. Ferroelastic toughening can also increase the fracture resistance of piezoceramics[4] , but it is not universally applicable and the toughening response varies across different systems. Making piezoceramic composites with polymers is considered a promising approach, offering enhanced toughness and compatibility with 3D printing.[26,27] However, the resulting composites exhibit low overall mechanical strength and limited polarization.[26,27] While suitable for sensing applications, their performance remains insufficient for energy storage/conversion devices. [28] In the context of electromechanical fatigue resistance, the presence of texture has been demonstrated as beneficial[29], but there are no studies to date that have addressed crack suppression through enhanced mechanical properties as a primary strategy for increasing lifetimes. Overall, these studies indicate a potential inverse relationship between piezoelectric and mechanical performance, where the addition of a reinforcing phase would increase toughness but be detrimental to the piezoelectric response.

The structure of natural materials has inspired researchers in recent past decades to use carefully designed microstructures that can break free from trade-offs inscribed into a material's composition.[30] While there are numerous examples of delicate microstructures to be reproduced, such as the impact-resistant hammer of the Mantis shrimp[31,32] or the open lattice structure of the starfish[33], the brick-and-mortar structure of nacre is the one associated with the highest toughness and mechanical durability.[30,34,35] Unfortunately, the toughness developed in nacre is based on the mortar being weaker than the brick. [36] This design rule makes it challenging to transfer to piezoceramics as most compositions present toughness lower than 1.5 $MPa.m^{1/2}$, which is on par with the toughness of the glasses typically used as the mortar phase in nacre-like composites.[30] Therefore, from a mechanical perspective, investigating the mechanical behaviour for such low-toughness materials with bioinspired nacre-like structures may inform broader toughening strategies for other fragile functional materials and advance our understanding of strategies to fabricate high performance, resilient materials.

Here, we propose a broadly applicable solution to simultaneously address the challenges and the open questions highlighted above by making a lead-free piezoceramic with a microstructure inspired by natural nacre. $Bi_{0.5}Na_{0.5}TiO_3$ (BNT) platelets, a promising piezoelectric composition[1,37], serve as the brick phase, while 0-5 vol% of colloidal silica functions as the mortar phase. Using a low-intensity magnetic field and a simple, safe, colloidal process, we produce bulk piezoceramics with aligned bricks. The final BNT-silica composites present a brick-and-mortar microstructure (B&M BNT) with a textured crystal structure. These materials provide piezoelectric performances close to the best BNT in the literature[1,38,39] but a 1.6- to 2-fold increase in toughness and up to a 2- to 3-fold increase in strength, reaching values comparable to structural ceramics.[30] We further demonstrate that this mechanical improvement translates into a significant 10 to 15-fold increase in ferroelectric and electromechanical fatigue resistance but also an increase in energy conversion capability, confirming the practical improvements our solution can make. We established that these improvements find their origin in both the formation of anisotropic grain morphology and the presence of residual stress fields, going beyond the strategy observed in natural nacre. Since all the enhancement mechanisms are rooted in microstructural designs, this strategy applies to other piezoelectric ceramics that can be processed into platelet form. Therefore, we believe that the multiscale-design nacre-like BNT can serve as a prototype for bioinspired nacre-like high-performance piezoelectric ceramics.

## Main text

We produced bulk BNT samples with both a high $<001>_{pc}$ texture and a brick-and-mortar microstructure through a low-intensity magnetic field-assisted colloidal process and the addition of a silica mortar. Details about low-intensity magnetic field-assisted bricks alignment can be found in the literature, including the effect of superparamagnetic nanoparticles amount and properties on the alignment quality[40,41]. The first step was to fabricate BNT bricks (Figure 1a), using a molten salt synthesis, from $Bi_4Ti_3O_{12}$ (BiT) precursors and topochemical microcrystal conversion. The bricks obtained were phase pure (Figure S1), with a length and thickness of around 5 - 11 µm and 0.5 - 1 µm, respectively (Figure 1b and Figure S2). The room temperature $R3c$ rhombohedral crystal structure of BNT originates from a high temperature $Pm\bar{3}m$ cubic structure, with the $<001>$ crystallographic direction of the high temperature cubic structure corresponding to the $<10\bar{2}>$ in the rhombohedral structure at room temperature (Figure 1a). Following literature convention[42,43], we also use $<001>_{pc}$ as the pseudo-cubic notation equivalent to the $<10\bar{2}>$ rhombohedral direction. Each platelet behaves as a monocrystal in electron backscatter diffraction (EBSD) (Figure S1), with the $<10\bar{2}>$ axis orientated in the brick thickness direction. After synthesis, bricks were functionalised with both superparamagnetic iron oxide nanoparticles and colloidal silica using heterocoagulation in water (Zeta-potential in Table S1 and energy dispersive x-ray spectroscopy (EDX) in Figure S3 to make them responsive to low intensity magnetic fields and ensure the presence of silica to form a mortar during sintering (Figure 1a, b and S4). Silica was chosen as it is conventionally used as a liquid phase sintering aid for structural ceramics and has good dielectric properties, including low loss and reasonable resistance to high electric fields. Functional ceramics pose an issue when choosing suitable mortars, as they are more brittle and softer than structural ceramics. While silica is used as a weak mortar with stronger bricks (alumina) [30,34], amorphous silica at this scale has similar structural properties (Young's modulus and fracture toughness) to BNT, so we would expect different strengthening mechanisms acting in these brick-and-mortar piezoceramics.

Our strategy of magnetic field-assisted processing uses safe and nontoxic components, whilst being easily scalable to bulk and 3D shape parts compared to tape casting. With the functionalised bricks ready to be assembled, we produced a dispersed slurry that was cast into a mould with a low intensity rotating magnetic field to align the bricks during their sedimentation. These were then consolidated into a green body as the solvent evaporated (Figure 1c and Figure S5). The green bodies were then sintered under pressure and with a fast heating rate using a Spark Plasma Sintering (SPS) furnace. A reference sample was prepared using a well-established oxide mixture and pressureless sintering. Both the magnetic-field assisted processing and SPS can introduce a texture. We confirmed that the magnetic-field assisted processing is responsible for the alignment as the texture quality of a 0 vol% $SiO_2$ sample, quantified by the Lotgering factor (as shown in Table S3), decreased slightly from 95.7% (green body) to 82.3% after SPS.

Next, we study the effect of both the processing with brick alignment and the presence of silica mortar on the microstructure obtained using a combination of EBSD and x-ray diffraction (XRD). The reference BNT presented a microstructure with an equiaxed grain morphology and 1 µm ± 0.4 µm grain size. No texture is observed from the EBSD and XRD patterns (Figure 1d and g). The magnetic field-assisted processing introduces a high degree of texture in all the samples, quantified by both the pole figure from EBSD (Figure 1e and f) and a Lotgering factor above 82% (Table S3). This texture is slightly lower than what can be obtained by tape casting and template grain growth[16–18], but higher than what is typically obtained with BNT[44]. In addition, our method has fewer limitations in sample size and brick orientation. A small amount of quartz and coesite, crystalline phases of silica, are visible with a peak at 28.64° and 29.53° (Figure 1g). Even with the high heating rate and pressure, the textured sample with no silica has an equiaxed microstructure with large 6 µm ± 3 µm grains and an aspect ratio around 1 (Figure 1h). This grain morphology is the most stable due to the low anisotropy of the cubic phase at high temperature. By adding a small amount of silica, the brick-and-mortar microstructure present before sintering is retained (Figure 1f). For B&M BNT, the higher silica content leads to a finer and more anisotropic final microstructure, with a final aspect ratio reaching 4 ± 2 with a silica amount of 5 vol% (Figure 1h). Microstructures of all samples are shown in Figure S6. Grain size and morphology are dictated by both thermodynamic and kinetic effects. Samples made from isotropic particles with 1 vol% silica, sintered by SPS at a same temperature and pressure, present (an equiaxed grain morphology (Figure S7). The brick-and-mortar structures obtained are thus kinetically stable, implying that the presence of silica slows down the diffusion of BNT, stopping it from reaching a more stable grain morphology. As such, we have designed a simple and robust process to obtain brick-and-mortar BNT (B&M BNT) with a high texture quality and established simultaneously the importance of silica to maintain the anisotropy in the grains.

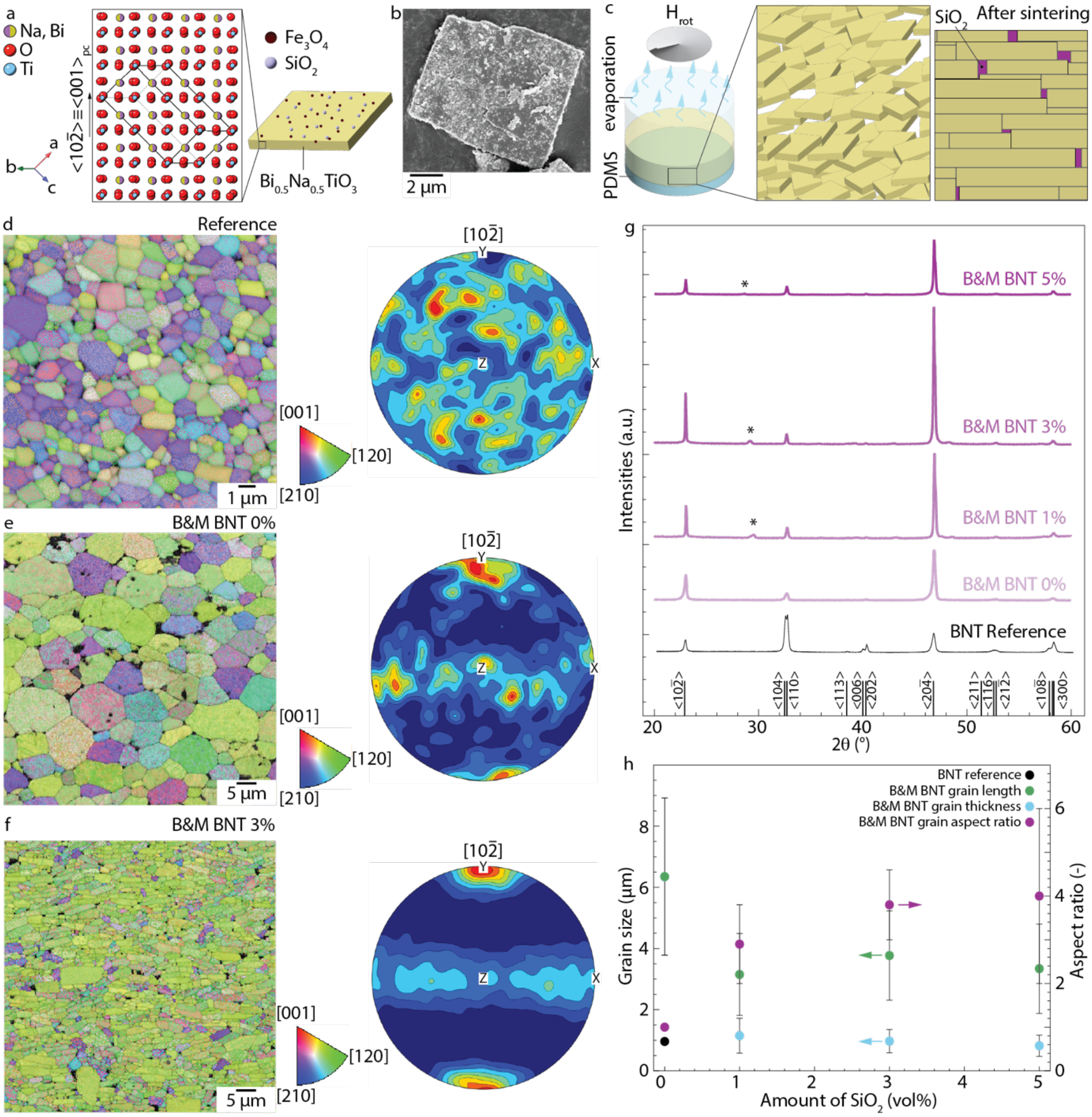

**Figure 1. Fabrication and microstructure of Brick-and-mortar (B&M) BNT.** (a) Link between the BNT crystal structure and brick shape, showing the equivalence between $<10\bar{2}>$ and $<001>_{pc}$ crystallographic direction. (b) SEM of BNT platelets coated with superparamagnetic iron oxide nanoparticles (SPIONS) and nano-$SiO_2$. (c) Magnetic-assisted assembly and drying process, built on a Polydimethylsiloxane (PDMS) substrate. The platelets are aligned parallel to the magnetic field rotation plane in ethanol, then undergo sedimentation, drying, and pressure-assisted rapid sintering. EBSD maps (Inverse Pole Figure along Y direction) for BNT reference (d), B&M BNT 0 vol% $SiO_2$ (e) and B&M BNT 3 vol% $SiO_2$ (f) cross section and pole figure of the $[10\bar{2}]$ plane for oriented platelets. (g) X-ray diffraction patterns for all samples, coesite or quartz silica mortar phases are indicated by "*". (h) Evolution of the grain size, length, thickness and aspect ratio with the volume fraction of silica in BNT. All grain dimensions have been measured by fitting ellipses on 2D cross-section images.

The mechanical properties of B&M BNT are significantly improved compared with those obtained with the reference BNT microstructure, which we found to be dictated by the size and spatial distribution of silica. The flexural strength measurements of the different compositions of BNT were made using three-point bending setup; 4-11 samples for each composition were tested with comparable relative densities of around 97% ~ 99%. The behaviour is linear elastic brittle, confirmed by fracture toughness tests showing no stable crack growth (Figure S8). Ferroelastic materials can demonstrate non-linear mechanical behaviour due to domain re-orientation and thus strain under stress. However, in the case of pure BNT, the poling stress or coercive stress has been measured around 306-371 MPa[39], which is 1.04 to 1.26-fold higher than the highest strength measured here for the 3 vol% $SiO_2$ sample. We thus chose to neglect the ferroelastic effect in the quantification of the strength as a first approximation. The high poling stress also points to a limited ferroelastic toughening in BNT-based composition which has been confirmed experimentally, even in compositions with a more relaxor-behaviour[45]. While ferroelastic toughening should be present, its effect will be limited and present in all our compositions, allowing us to focus our explanation on the effect of microstructure and presence of silica. The introduction of 3 vol% silica and brick-and-mortar architecture improves the strength and toughness of BNT by almost 3-fold (from 104 to 294 MPa) and 2-fold (from 0.9 to 1.8 $MPa \cdot m^{0.5}$), respectively (Figure 2a and b), with respect to the reference BNT with the equiaxed grains. The strengths measured are up to 294 MPa (Figure 2b) and therefore on par with brittle, conventionally produced structural ceramics such as alumina and silicon carbides[46]. Fracture toughness and strength follow similar trends across all compositions. In linear elastic brittle solids, both quantities are linked by the relationship $K_{IC} = \sigma Y \sqrt{a}$ with $K_{IC}$ the fracture toughness in mode I, where $\sigma$ is the strength, $Y$ is the stress concentration factor, and $a$ is the critical defect size. This relationship suggests that increasing the amount of silica both increases the toughness and reduces the size of the critical defect. The increase in strength and toughness comes from both the presence of silica and brick-and-mortar architecture: the reference sample and B&M BNT of 0 vol% $SiO_2$ have the same toughness, and adding 1 vol% $SiO_2$ in the reference BNT sample raises the strength and toughness by only 22% and 20%, respectively, while in B&M BNT with 1 vol% $SiO_2$ gives 61% and 117% improvements in strength and toughness, respectively. The presence of silica and the brick-and-mortar structure not only increases the strength and toughness, but also the Young's modulus and strain at failure (Figure 2c and d), reaching a strain 142% higher compared to the reference BNT.

This all-round structural improvement suggests the introduction of toughening mechanisms. However, in all the other fully ceramic brick-and-mortar compositions, large-scale deflection and local crack branching lead to stable crack growth.[30,34] In B&M BNT, the silica has similar toughness and stiffness as BNT[47,48], the fracture remains brittle, and there is an optimum silica amount at 3 vol%, after which the properties decrease again (Figure 2a and b).

The toughening mechanisms arise from the sample microstructure and presence of silica, and we used EDX maps and transmission electron microscopy (TEM) to observe the localisation and size of the silica. EDX (Figure 2e-h) shows that the silica is present almost exclusively as pockets at the edge of the bricks, with an average size of these silica pockets increasing as more is added, from 0.9 μm ± 0.5 μm (1 vol% $SiO_2$) continuously increasing up to 2 μm ± 0.8 μm (5 vol% $SiO_2$) (Figure 2f). The distance between the silica pockets is estimated as the average space around a single silica pocket and calculated from the number of silica pockets in a given area and their area fraction. The separation distance first decreases from 4 μm to 3.2 μm, then increases to 5 μm as the silica content increases from 1 vol% to 5 vol% (Figure 2f). This indicates that as the amount of silica added increases, the degree of aggregation also increases. We attribute this to the more readily available silica presence on the bricks as the content increases, so with similar diffusion time, the silica can form larger pockets at the edge of the bricks. The TEM analysis confirms this observation, with the added presence of silica film of around 9 nm between BNT bricks close to the pockets (Figure 2g). From the TEM observation, the silica film thickness and amount do not change significantly when the silica content increases from 1 to 5 vol% (Figure 2h and Figure S9). Therefore, the $K_{IC}$ and strength value variations with silica volume fraction cannot be explained by the presence of silica at the grain boundary. The iron oxide nanoparticles also aggregate during sintering, forming pockets at the bricks edge, (Figure 2g and S10) and their presence is sparse due to the low volume fraction used.

Assembling all these observations, we can propose a new mechanism by which the silica spatial repartition can influence the structural properties. Silica has similar structural properties to BNT but has 2.4- to 25- times lower coefficient of thermal expansion (CTE) of $4.79 \times 10^{-6}$ /K (< 150 °C)[49] and $1.03 \times 10^{-5}$ /K (> 150 °C)[49] compared to $0.4 \times 10^{-6}$ ~ $4.1 \times 10^{-6}$ /K (> 300 °C) for silica[50–53]. The concept of placing inclusions in compression within a matrix under tension was applied more than two decades ago to form nanocomposites, for instance, SiC within alumina[54,55]. However, this strategy did not increase toughness, only strength. This effect can be observed by comparing the reference BNT with 0 and 1 vol% $SiO_2$ (Figure 2a and b), leading to a minor 20% strengthening and 22% toughening, compared to the ones observed in B&M. Using inclusions was later used in piezoceramics, with ZnO in BNT-BT[56]. The addition of ZnO inclusions lead to an increased resistance to thermal depoling, which was associated with zinc-doping of the base materials[57]. In our B&M composition, silica pockets will be under compression after cooling due to the thermal expansion mismatch and will put adjacent bricks under tension, which we verified using XRD (Figure S11). A simple isotropic approximation of the magnitude of residual stress can be obtained using classical models[58] and leads to a value ranging from 264 to 256 MPa as the silica content increase from 1 to 5 vol%. This value is in the same order of magnitude with the increase in strength measured (Figure 2b), with the strength increasing by around 150 MPa as the silica content increasing from 0 to 3 vol%. However, in our case, the spacing and position of silica pockets are dictated by the BNT brick and thus are anisotropically spaced, creating anisotropic residual stress fields that interact with a crack as it initiates. Toughening has been described theoretically for inclusions generating compressive stresses in their surroundings, where the high number and low distance between inclusions increase the strength and toughness of the composite, for instance, using Eshelby[59] or Mori-Tanaka[60] models or more recently using numerical models[61]. While in these models the toughening comes from compressive stresses in the matrix, in our case, it comes from compressive stress in the silica pockets and tensile stresses in the bricks.

Our assumption is that the anisotropy and the spatial variation of the residual stress is responsible for the increase in toughness by modifying the stress intensity field, however, we would need to access the precise spatial variation of residual stresses field to further model this effect. We can however quantify the presence of anisotropic and locally varying residual field using Raman spectroscopy. We mapped the shift of the Raman peak attributed to the $TiO_6$ octahedra with $A_1$ symmetry, centred around 275 $cm^{-1}$ in stress-free crystals[62] and around 276 $cm^{-1}$ in our reference BNT sample (Fig. S12), in the samples with 0 and 3 vol% of $SiO_2$ along and perpendicular to $< 001 >_{pc}$ (Fig. 2 i-l). These results reveal first an increase in the area showing a peak shift and thus residual stresses with the increase in silica content, with an average peak position of 274.7 $cm^{-1}$ parallel (Figure 2i) and 274.4 $cm^{-1}$ (Figure 2j) perpendicular to $< 001 >_{pc}$ and 273.5 $cm^{-1}$ parallel (Figure 2k) and 272.2 $cm^{-1}$ perpendicular (Figure 2l) to $< 001 >_{pc}$ for 0 and 3 vol% $SiO_2$, respectively. Only down shifts are observed for this peak in all samples, confirming the presence of only one sign of residual stresses in BNT, while the silica should be stressed oppositely but is not detected in this mapping. The mapping further highlights the anisotropy of the residual stress field in B&M samples, with the average peak position being 0.48% lower in the direction perpendicular (Figure 2l) to $< 001 >_{pc}$ than parallel (Figure 2k). Finally, we can directly observe high spatial variation of the residual stresses in the 3 vol% $SiO_2$ samples, with peak shifting from 274.9 $cm^{-1}$ down to 266.4 $cm^{-1}$ over distances of a few microns (Figure 2k, l). We also confirmed that at 5 vol% of SiO2 the overall peak shift is lower (Figure S12) than with 3 vol%, confirming the trend observed with the strength and toughness.

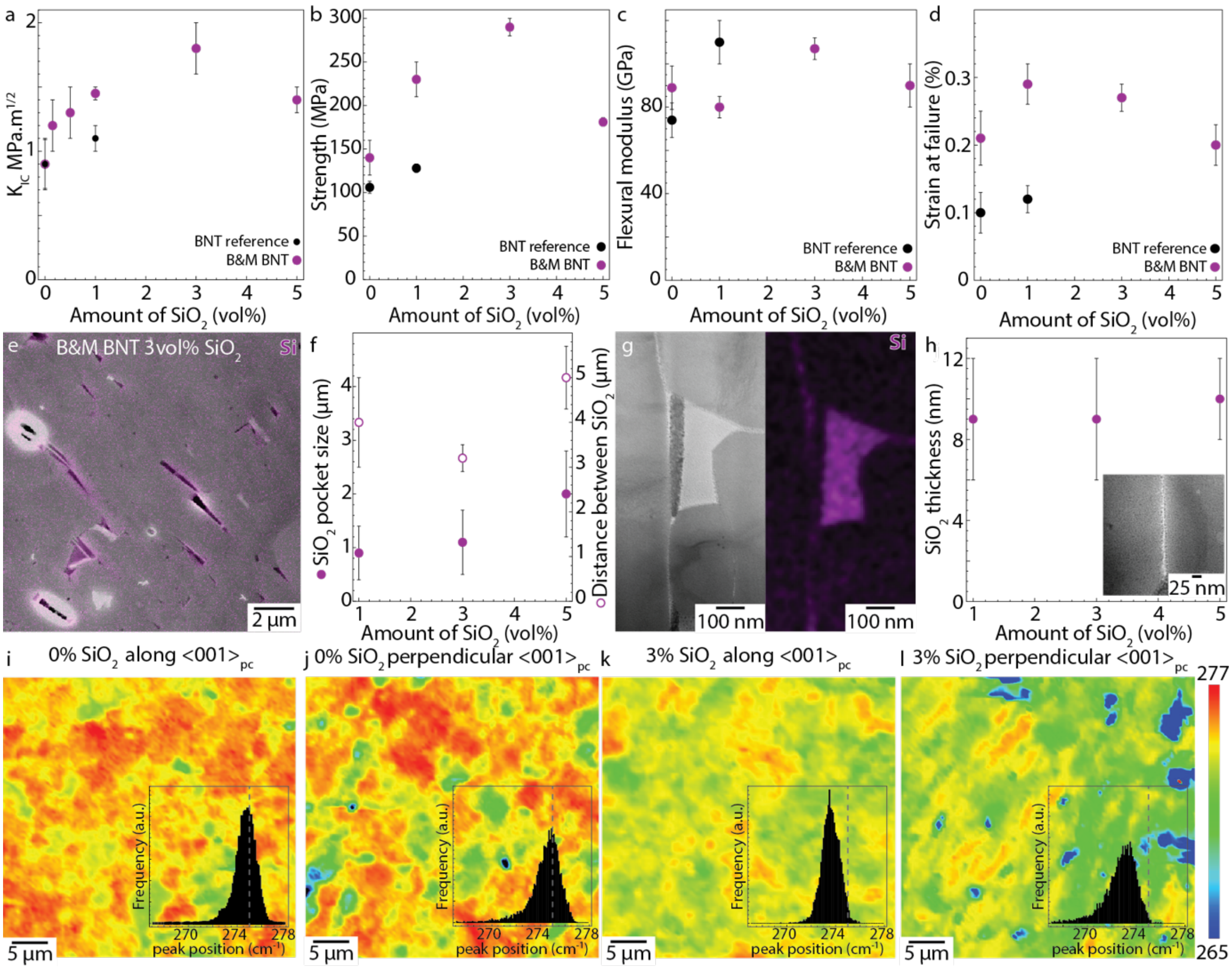

**Figure 2. Mechanical properties of B&M BNT with different amount of silica.** Evolution of the fracture toughness $K_{IC}$ measured in single-edge-notch bending (SENB) (a), the flexural strength (b), the Young's modulus (c) and strain at failure (d) as a function of silica volume fraction for reference and B&M BNT samples. (e) Overlay of SEM and map of Si element from EDX maps for B&M BNT with 3 vol% $SiO_2$. (f) Average size and distance of $SiO_2$ pockets from EDS maps as a function of $SiO_2$ volume fraction. (g) STEM Bright field image of a $SiO_2$ pocket and corresponding STEM EDX Si map for B&M BNT 3 vol% $SiO_2$. (h) Evolution of the thickness of $SiO_2$ at the grain boundary in B&M BNT from TEM images as function of $SiO_2$ volume fraction. Inset: TEM image of a brick boundary with $SiO_2$ in B&M BNT 3 vol% $SiO_2$. Map of BNT Raman fitted peak position associated to the $TiO_6$ octahedra of A1 symmetry[62,63] for 0 vol% $SiO_2$ along (i) and perpendicular (j) <001>$_{pc}$ and for 3 vol% $SiO_2$ along (k) and perpendicular (l) <001>$_{pc}$. All the maps are represented with the same colour scale present on the right of the maps (unit $cm^{-1}$). Inset: histograms of the peak shift for each map, dotted line represents the peak position in stress-free sample[62].

The presence of both silica pockets and the brick-and-mortar microstructures increases the strength and toughness of B&M BNT by introducing an anisotropic and locally varying residual stress fields. While this improvement makes these compositions compete with some structural ceramics, their functional properties need to remain as high as the reference composition to be useful piezoceramics.

We now demonstrate that the strong and tough B&M-BNT are also high-performing piezoceramics compared with the reference BNT due to the texture and brick-and-mortar microstructure. Polarisation versus electric field (P-E loops) and strain versus electric field (S-E loops) as a function of electric field cycling are shown in Figure 3. For the reference BNT, 7.5 kV/mm was the highest field it was possible to apply before dielectric breakdown occurred. The measured values fall within the expected range for BNT-based ceramics.[1,38] Notably, higher polarisation was observed in the B&M BNT of 0 vol% and 5 vol% $SiO_2$ samples under higher fields. This enhancement is attributed to a combined effect of 1) texture, which provides a favourable crystal orientation $< 10\overline{2} >$ relative to the applied electric field direction, and 2) these samples having lower residual tensile stresses compared to samples with an intermediate amount of $SiO_2$. It should also be kept in mind that the P-E loops of the reference BNT could only be recorded at a lower electric field than the other samples, which contributes to its lower $P^r$. The $E_c$ is also lower in all the samples with silica compared to the 0 vol% $SiO_2$ sample. This can be an effect of the smaller grains in the B&M structure[64] or the presence of residual stresses.

Similar trends are observed in the S-E loops, with B&M BNT with 0 vol% and 5 vol% $SiO_2$ exhibiting the highest strain compared to intermediate silica containing B&M BNT and reference BNT ceramics. In the unpoled state, the largest strain is obtained for the 5 vol% containing ceramics (112 pm/V) followed by the 1 vol% and 0 vol% (100 pm/V), corresponding to improvements of 84% and 64%, respectively, compared to the reference BNT ceramic (measured at 7.5 kV/mm). Poling does not change the electromechanical response in the textured samples, while the reference sample showed a higher response at the same applied electric field after poling, as shown in Figure 3c at 7.5 kV/mm.

The value of $d_{33}^*$ shown in Figure 3d, calculated from the maximum strain-field response before breakdown, is highest for the samples before poling, except for the reference BNT ceramic. While it is possible that excessive poling decreases the $d_{33}^*$ due to poling-induced irreversible long-range ferroelectric order formation [65,66], the $d_{33}^*$ of the poled textured samples are calculated at a lower electric field amplitude than the unpoled, which usually lowers the $d_{33}^*$.[67] The highest $d_{33}^*$ (111 pm/V) remains in the 5 vol% silica composite even after poling. The values of $d_{33}$ after poling are again higher for 0 vol% and 5 vol% samples, with 91 pC/N and 95 pC/N, which is a 25% and 30% increase compared to the reference sample, respectively. The 1 vol% samples also increased by 14% up to 83 pC/N.

The correlation between $d_{33}$ and Vickers hardness is highlighted in Figure 3e, with both values displaying contrary trends. Higher hardness corresponds to a higher residual stress field generated by the silica pockets. This residual stress field interacts with and pins domain walls, reducing non-180° domain switching and limiting the piezoelectric response. This correlation could be explained by the anisotropic dispersion of silica pockets toughening and hardening the composite matrix, but also affecting the mobility of domains during polarisation. While the silica pockets might limit domain mobility, all samples, both conventional and brick-and-mortar architectures could still undergo ferroelectric switching and display reasonable piezoelectric response.

The performance of piezoceramics in, e.g., sensing and energy harvesting applications can be expressed with more comprehensive indexes, such as $g_{33} = d_{33}/\varepsilon_{33}^T$ = 18.3×10$^{-3}$ Vm/N and $FoM = d_{33}g_{33}$ = 1.7 pm$^2$/N (Figure 3f). Both values further highlight the trends found with the P-E and S-E loops, with some differences coming from the changes in relative permittivity with the composition, texture, and brick-and-mortar microstructure (Figure S13). In this case, both 1 vol% and 5 vol% have the highest $g_{33}$ and $FoM$ values, suggesting their potential use as sensors in high-stress environments or as energy harvesters.

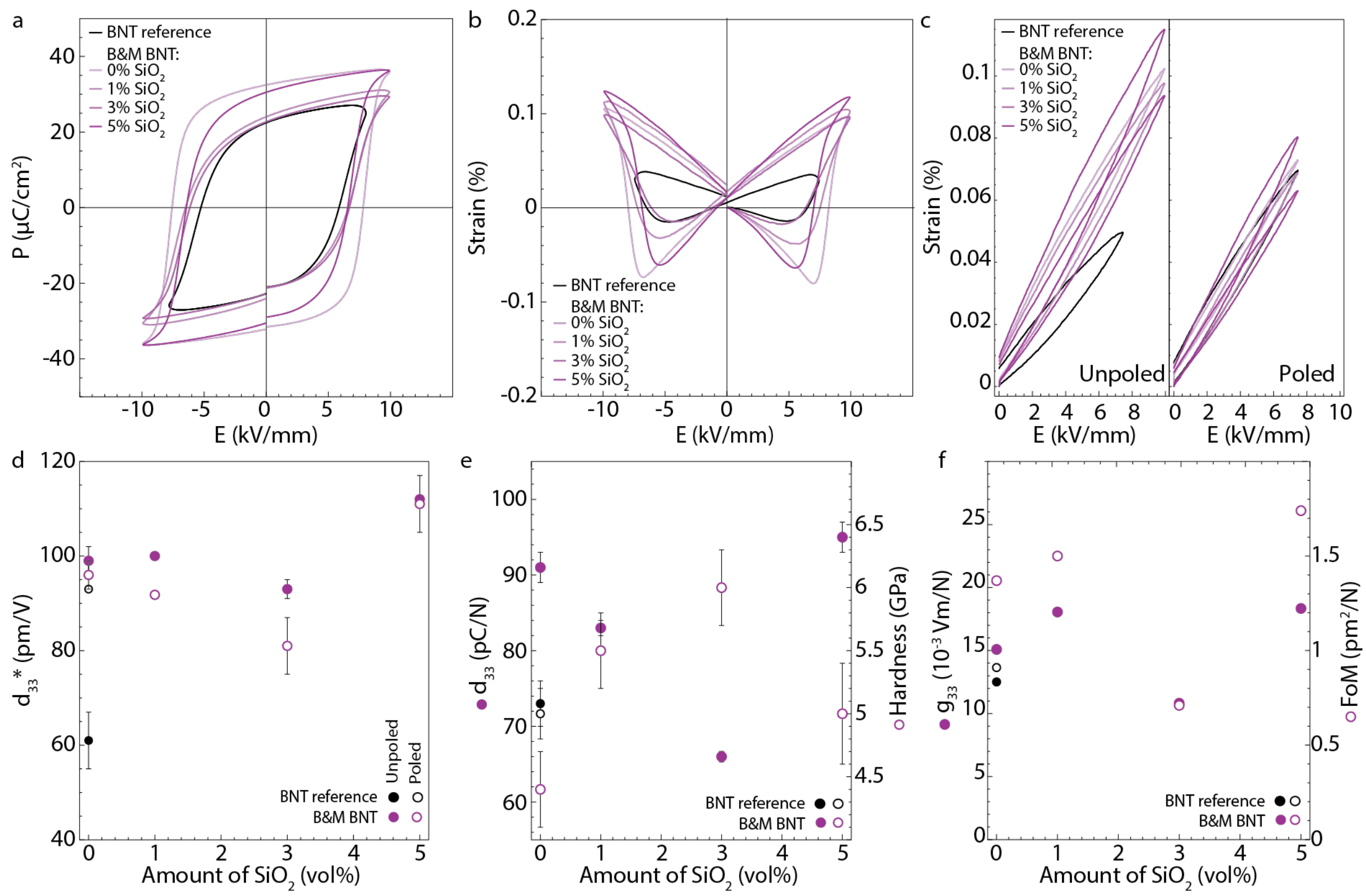

**Figure 3. Piezoelectric and ferroelectric performance of B&M BNT ceramics and BNT reference.** Ferroelectric P-E hysteresis loops (a) and S-E butterfly curves (b) at 10Hz for BNT reference and B&M BNT 0-5 vol% $SiO_2$. (c) Unipolar strain-electric field (S-E). (d) Field-induced piezoelectric strain $d_{33}^*$ evolution (calculated from data in (c)) as a function of silica volume fraction for reference and B&M BNT samples. (e) Evolution of piezoelectric constant $d_{33}$ and Vickers hardness as a function of silica volume fraction for reference and B&M BNT samples. (f) Evolution of the piezoelectric voltage coefficient $g_{33} = d_{33}/\varepsilon_{33}^T$ and piezoelectric Figure of Merit $FoM = d_{33}g_{33}$ value for all samples at 1kHz as a function of silica volume fraction for reference and B&M BNT samples.

Whereas the optimum amount of silica is different for mechanical and piezoelectric properties, we have confirmed that the presence of silica and a brick-and-mortar microstructure also increases the piezoelectric and ferroelectric properties of the materials. We also found that a modulation of the local residual stress fields induced by thermal expansion mismatch between the BNT platelets phase and silica mortar phase affects both the hardness and overall piezoelectric and ferroelectric properties.

We now explore how this new combination of mechanical and piezoelectric properties affects the material durability and performance. The improvement in mechanical properties observed along with the good piezoelectric properties of our B&M-BNT translates into high-performing and highly durable materials. We can first test how the dual increase in mechanical and piezoelectric properties translates into durability through cyclic fatigue testing under bipolar electric field cycles at 1000 Hz and measured at 10 Hz (Figure 4a). The polarisation reached for each sample is in line with what was obtained previously, and the maximum cycling number was capped at $10^6$ to limit the test duration. The presence of texture is beneficial to the fatigue lifetime, increasing the number of cycles it takes for the remanent polarisation to reach 80% of its initial value $P_0^r$ by 10 to 15-fold with respect to the reference (Figure 4b). In addition, the presence of silica and the brick-and-mortar microstructure not only increases the number of cycles by 10 compared with the reference sample but also decreases the polarisation fluctuations (Figure 4a). The presence of silica and the designed microstructure that induces a residual stress field influencing the domain rotation as well as the higher fracture toughness which delays microcrack initiation limits the cumulative damage from each cycle and ensures the durability of the material. This is demonstrated from 2/3 of the 1 vol%, 3/4 of the 3 vol%, and all 3 of the 5 vol% samples tested for fatigue surviving $10^6$ cycles under 7.5 kV/mm without failing. The cycles to failure for B&M BNT are in the range of what is obtained with BNT-based materials tested in bipolar fatigue in the literature[68], however the different cycling frequency used prevents direct comparison on the polarisation decrease and we limit our analysis to compare the materials fabricated for this study.

One clear trend is that higher overall mechanical strength and toughness correlate positively with fatigue resistance, as tougher ceramics are more resistant to crack initiation under long-term ferroelectric fatigue. This is even more impressive as we textured along the $< 10\bar{2} >$ direction the samples, the direction that leads to the highest piezoelectric response and so the one leading to the highest electromechanical strain. Additionally, previous literature[69–71] suggests that externally applied compressive prestress improves fatigue resistance but can reduce unipolar polarization—especially in multilayer actuators (MLAs). However, unlike MLAs, the matrix in our material experiences internal tensile stress rather than compressive preload, which may facilitate domain switching without sacrificing polarization.

While the interaction of the residual stress field and cyclic polarisation efficiency and durability require further study to be understood fully, we can already test the practical electrical output of our materials, which is of significance for transducer applications, e.g. sensing and energy harvesting. Using an electromagnetic shaker capable of providing one-dimensional reciprocating motion $u(f)$ at variable frequencies and a fixed weighted cantilever setup, we recorded the voltage generated by the strain applied to the sample with B&M BNT of 1 vol% $SiO_2$ compared with a reference BNT. Given the similarity in resonance frequency, we can assume that both samples experience similar strains. The peak-to-peak open-circuit voltage ($V_{p-p}$) of the B&M-BNT 1 vol% $SiO_2$ cantilever beam generated at its resonance frequency (22 Hz) was almost 1.5 times higher than that of the reference sample. Similar advantages were consistently observed over the whole frequency range of 2 to 26 Hz (Figure 4c). While in this configuration the voltage generated is related to the $d_{31}$ coefficient, the conclusions we draw so far on $d_{33}$ are confirmed, with a higher voltage generated in the B&M BNT. Not only does the B&M BNT provide a 46% higher voltage output compared with the reference sample, but it also increases the durability of the harvester. This is demonstrated by comparing the relative decrease in voltage of after 180k cycles of oscillation. which was 30% for the B&M BNT composite with 1 vol% $SiO_2$ compared with 50% with the reference samples (Figure 4d). Our B&M BNT samples thus harvests more overall energy for more cycles, compounding the gain over the whole lifetime of the materials in use.

As a final step, we can compare the overall performances of the different compositions of B&M BNT (Figure 4e). All our B&M-BNT compositions present performances higher than the reference BNT, apart from the $d_{33}$ of the 3 vol%. Adjusting the amount of silica and thus altering residual stresses within the B&M BNT provides an effective strategy to design materials for real engineering application requirements, for instance, either emphasising more the piezoelectric performance and durability with the 5 vol% materials, or the mechanical performance with the 3 vol% composites, with 1 vol% silica concentration representing a compromise between the two. Finally, we can put our strategy in context with other available piezoelectric ceramics (Figure 4f). If we represent the strength versus the direct piezoelectric coefficient $d_{33}$ for a wide range of compositions, we observe that higher $d_{33}$ generally translates into lower mechanical properties. While mechanical properties are generally not considered as critical for these functional materials as their electromechanical coupling, they are effectively linked with the durability and performance of the materials in use as we have seen with our study. Our strategy to use silica pockets and a brick-and-mortar microstructure manages to increase the mechanical performance of BNT to reach the strength of structural ceramics, without compromising its functional properties, reaching a new area in this map. We confirm that these strength improvements originate from an increase in toughness, itself originating from the presence of anisotropic stress fields, which translates into easier domain switching, orders of magnitude longer lifetime and higher energy conversion capability. This strategy is not limited to BNT composition and can potentially be applied to any composition that can be synthesised as monocrystalline bricks as silica has one of the lowest CTEs of any oxide material. We confirm the broad applicability of this bioinspired hierarchy strategy by showing the increase in strength and toughness while retaining similar or even higher$d_{33}$ in BiT and $0.94Bi_{0.5}Na_{0.5}TiO_3$-$0.06BaTiO_3$ and higher toughness for $BaTiO_3$ using a similar processing method and adding various amount of silica (Table S4). Going further, we envision that the interplay with texture, microstructure, residual stress field and structural/functional properties relationship is a field of research that could bear more fruit within this class of technologically critical material.

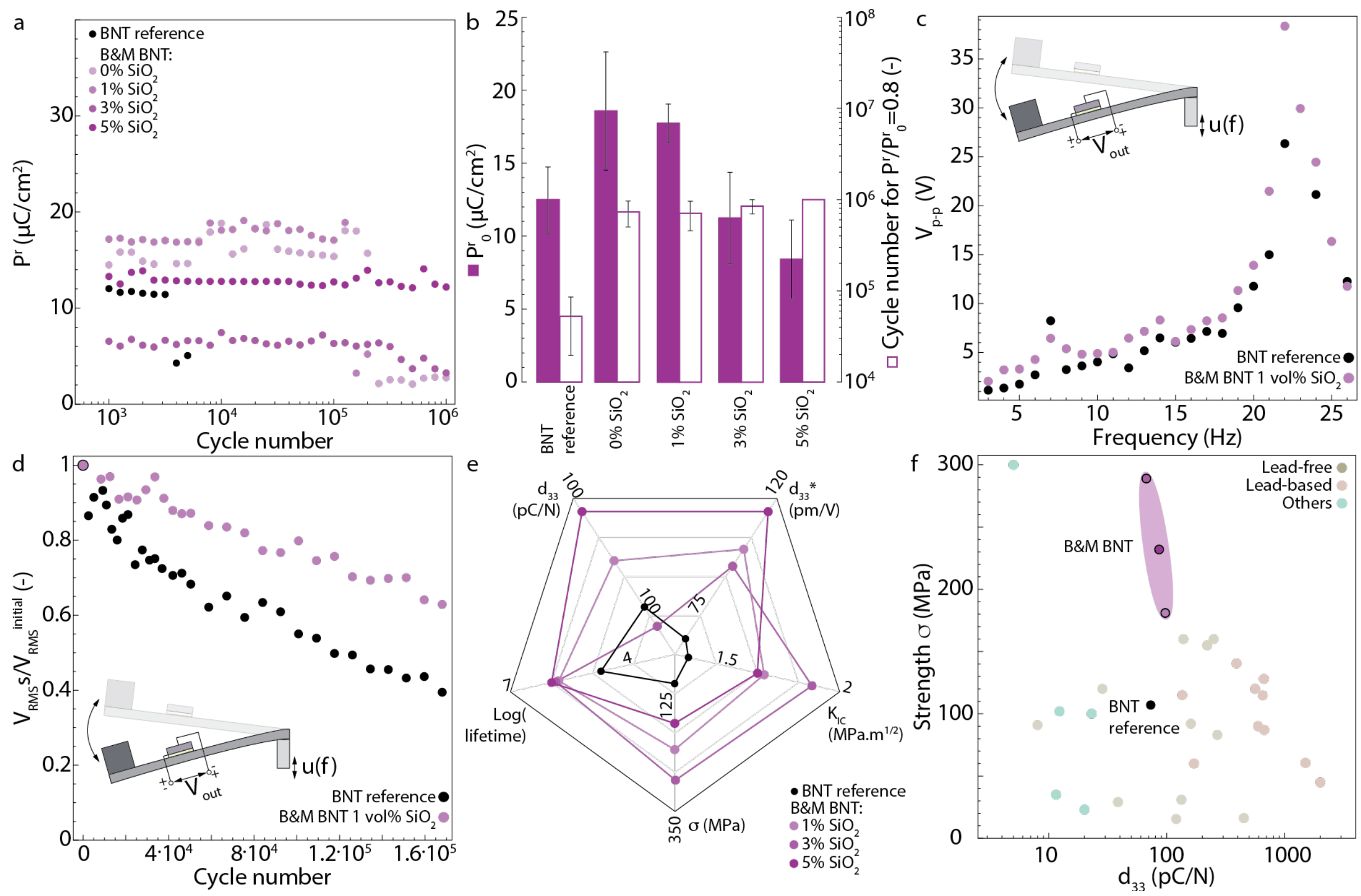

**Figure 4. Mutually beneficial structural and functional improvements lead to performant and durable piezoceramics for energy harvesting.** (a) Bipolar P-E loop fatigue performance with remnant positive polarization as a function of cycle number under 7.5 kV/mm. (b) Initial polarisation $P_0^r$ and cycle number for which 80% of the positive polarisation remains for the different samples. Note, the 5 vol%. samples and some 1 and 3 vol% samples did not show such a polarisation decrease in more than $10^6$ cycles so this cycle value was used for these samples. (c) Frequency dependency of peak-to-peak open circuit $V_{p-p}$ of the cantilever beam-based transducer integrated, with the reference BNT and 1 vol% $SiO_2$ B&M BNT, with $u(f)$ in the set-up model schematic represents the bidirectional reciprocating motion signal of the variable-frequency output under a constant driving force. (d) Cycle dependency of root mean square open-circuit voltage ($V_{RMS}$) of the cantilever beam transducer at an oscillation frequency of 14Hz, which was normalised by the initial value of the $V_{RMS}$, and the $u(f)$ represents the bidirectional reciprocating motion signal at a fixed output frequency of 14 Hz under constant driving force. (e) Summary of the mechanical performance, piezoelectric performance, durability and energy harvester ability for B&M composite and BNT reference. (f) Strength as a function of $d_{33}$ coefficient for our B&M BNT composition, BNT references and data from the literature based on lead-based (PZT-based[72–75], PMN-PT[72], PMN-PZT[72], PZT composites[75–77]) and lead-free (KNN-based[78], BNT-based[1,79–82], BT-based[80], $Ba(Zr_{0.2}Ti_{0.8})O_3$-based[80,83], $BiFeO_3$-based[84,85], $Bi_4Ti_3O_{12}$-based[86–88], PVDF-based[27,89,90], other piezoceramics including ZnO and AlN polymer piezoceramic composites[91–94]) piezoelectric materials.

## Conclusion

In this research, we successfully implemented a $<10\bar{2}>$-($<001>_{pc}$-) textured multi-scale bioinspired brick-and-mortar architecture into lead-free piezoceramics, with $Bi_{0.5}Na_{0.5}TiO_3$ (BNT) as the brick and the amorphous silica as mortar (B&M BNT). Crystallographic texture was introduced alongside a brick-and-mortar grain morphology through the addition of silica to BNT platelets that were aligned through magnetically assisted slip casting followed by pressure-assisted sintering. The introductions of both this specific grain morphology and silica provide a 2- to 3-fold and from 1.6- to 2-fold improvement on the strength and toughness of the material, respectively. The toughening, based on anisotropic residual stress field, expands our toolbox of mechanisms and is especially suited for functional materials, where the 'bricks', i.e. the piezoceramic phase, are mechanically weak. For specific silica and BNT compositions, this increase in mechanical properties preserves the piezoelectric performances and, in certain cases, improves them compared with a non-textured sample. Because of this combination of toughening and piezoelectric performances, we explored the link between mechanical properties and piezoelectric-effect-driven fatigue resistance. Our brick-and-mortar composites show a significant increase in durability. As sensors, the brick-and-mortar composition shows at least a 10-fold increase in ferroelectric fatigue resistance, with some compositions showing no change in polarisation after $10^6$ cycles, which is a 15-fold increase compared with the reference BNT. As energy transducers, they not only show a 46% increase in output voltage but also display an obviously slower degradation rate and longer lifetime compared with the reference BNT samples. Finally, this work establishes B&M BNT as a prototype for an advanced functional ceramic design strategy. As the enhancement mechanism arises entirely from microstructural engineering, this concept can be readily extended to a broad range of piezoelectric systems, offering a pathway toward overall enhancement of mechanical and piezoelectric performances across diverse ceramic compositions.

## Materials and Methods

Synthesis of $<001>_{pc}$-oriented $Bi_{0.5}Na_{0.5}TiO_3$ (BNT) brick powder and reference BNT equiaxed powder

The $< 10\bar{2} >$-oriented BNT bricks powders were fabricated by molten salt synthesis (MSS) and topochemical microcrystal conversion (TMC) inspired by previous studies[95–98]. First, $< 001 >$-oriented $Bi_4Ti_3O_{12}$ (BiT) platelets templates precursors were synthesized via MSS: $Bi_2O_3$ (Sigma-Aldrich, 99.9%, with 3 wt% excess amount used to ensure the Bi : Na ratio, as Bi evaporation during synthesis can influence the stoichiometry) and $TiO_2$ (Sigma-Aldrich, 99.8%) under molten salt mixture of NaCl (VWR, >99.5%) and KCl (Sigma-Aldrich, >99.0%) (molar ratio 1:1). The salt mixture and reagents were mixed in a weight ratio of 1:1 in ethanol and ball milled for 24 h. Next, the $< 1\bar{1}2 >$-oriented BNT platelets were fabricated based on precursor BiT templates by Topochemical Microcrystalline Conversion (TMC). The BiT were mixed with $Na_2CO_3$ (Sigma-Aldrich, >99.5%), $TiO_2$ (Sigma-Aldrich, 99.8%) in NaCl (VWR, >99.5%) in ethanol, with weight ratio of reagents to salt of 1:2. The mixture was heat treated in air at 800 °C for 3 h. The BiT and final BNT platelets were both washed out from mixture by deionized water. The XRD and SEM EBSD mapping figures for BNT platelets are shown in Figure S1 and the powder size / morphology distributions are shown in Figure S2. The reference equiaxed BNT powder was synthesized by conventional solid-state reaction from a mixture of $Bi_2O_3$ (Sigma-Aldrich, 99.9%, with 3 wt% excess amount used compared with stoichiometric amount, same to ensure the Bi : Na ratio and high resistivity), $TiO_2$ (Sigma-Aldrich, 99.8%) and $Na_2CO_3$ (Sigma-Aldrich, >99.5%) by ball milling for 24h in ethanol. The mixture was dried and calcinated under air at 800 ℃ for 3 h. The 1-step molten-salt-synthesized equiaxed BNT powder was also obtained by the same recipe for solid state, while in NaCl (VWR, >99.5%) in ethanol, with weight ratio of reagents to salt of 1:2, calcinated in air at 800 °C for 3 h.

Fabrication of brick-and-mortar BNT (B&M BNT) ceramics and reference samples

To fabricate B&M BNT ceramics, firstly, the BNT platelets were functionalized by superparamagnetic iron oxide nanoparticles (SPIONS, ferrofluid EMG605 from FerroTec) with 37.5 μl/g in a water-based suspension. The functionalized BNT platelets were coated with different amounts of nano-silica (1 vol%, 3 vol%, 5 vol%) by adding LUDOX® AS-50 colloidal silica in the water-based suspension system, stirring for 24 h. The powder SEM image is shown in Figure S4.

Next, for the green body texturing, a system composed of functionalized and silica-coated BNT platelets, ethanol and 5 wt% polyvinylpyrrolidone (binder) was used. The slurry was injected into a plastic cylinder mold (Ø3 cm × 5 cm). A 500 mT permanent neodymium magnet was rotated horizontally (NIBL 00954, N38H grade, Magnet Sales, rotation speed > 250 rpm) to align the platelets within this suspension system, which was positioned around 2 cm from the mold. The green body was dried in air at room temperature for 24 - 36 h. More information on the texturing procedures can be found in reference[35]. After debinding at 550 °C for 4 h, the green body was moved into the spark plasma sintering furnace (KCE®-FCT HP D 10-SD from FCT System) using a graphite mold (Ø3 cm), and was sintered under argon at 1030 °C for 5 min with a heating rate of 50 °C/min. The sintered B&M ceramics were annealed under 800 - 820 °C in air for 3 h to remove the carbon. For the normal reference BNT, the equiaxed powder was pressed into Ø3 cm pellets by uniaxial pressing under 100 MPa and cold isotropic press 250 MPa respectively, sintered by conventional sintering under 1100 °C for 2 h. For the equiaxed BNT with 1 vol% $SiO_2$, the 1-step molten-salt-synthesized equiaxed BNT powder was coated by LUDOX in water suspension system, then a similar ethanol-based suspension system was fabricated and poured into a plastic mold on gypsum and dried into a green body for 24 h. This green body was sintered by SPS under the same conditions as B&M BNT (disk-shaped pellets with 30mm in diameter and 3-5 mm in thickness).

Microstructure, morphology and texturing quality characterization

The existing phases and chemical compounds were measured by X-ray diffraction (XRD) via Bruker D2 Phaser. The Lotgering Factor[1] was used to evaluate the $\langle 1\bar{0}2\rangle = \langle 001\rangle_{pc}$ texturing quality, which was calculated based on XRD according to following equation:

$$F_{(00l)} = \frac{P - P_0}{1 - P_0} \quad (1)$$

$$P = \frac{\sum I_{(00l)}}{\sum I_{(hkl)}} \quad (2)$$

$$P_0 = \frac{\sum I_{0(00l)}}{\sum I_{0(hkl)}} \quad (3)$$

The scanning electron microscope (SEM) figures, energy-dispersive X-ray spectroscopy (EDX) mapping and electron backscatter diffraction (EBSD) pattern were detected by TFS Quanta ESEM and Zeiss Sigma300. The grain orientations and microstructure were analyzed by EBSD pole figure and inverse pole figure reindexed by MTEX[99]. EBSD data and pole figures were plotted based on these data. The powder size distribution and grain size distribution were quantitively analyzed by SEM and ImageJ[100] by fitting ellipses on segmented images to obtain the large and small diameters of the ellipses, which were taken as the thickness and length of the platelets. STEM-EDS data was obtained using a TEM JEOL STEM 2100F and was used to determine the morphology and local chemistry at the platelet-matrix interfaces. Raman spectra were acquired using a confocal Raman microscope (inVia, Renishaw, U.K.) with a 532 nm excitation laser. Raman mapping was performed with a step size of 400 nm over an area of 40 μm × 40 μm.

Mechanical testing

The strength and toughness were determined by in-situ three-point bending, with a span of S = 20 mm and measured on 4 - 11 samples for strength and 4-6 samples for toughness for each composition. For toughness, single-edge-notched beam (SENB) toughness test specimens were cut from the sintered pellets. The ceramics were machined into beams with a length of 30 - 20 mm, thickness of W = 3 - 5 mm and width of B = 3 - 4 mm (B<W). The notch length was between 0.4W < a < 0.6W, with width of 0.5 mm, and the notch tip was sharpened according to ASTM E1820[101]. The three-point bending fracture toughness was tested in a 10 kN universal testing machine (Z010, ZwickRoell) under constant displacement rate of 1 μm/s. For the calculation, the critical stress intensity factor $K_{IC}$ was calculated using ASTM E1820:

$$K_{IC} = \left[\frac{FS}{BW^{3/2}}\right] \times f\left(\frac{a}{W}\right) \quad (4)$$

where *F* is the applied force when the fracture occurs, and $f$ is given by:

$$f\left(\frac{a}{W}\right) = \frac{3(\frac{a}{W})^{\frac{1}{2}} \times [1.99 - \frac{a}{W}(1 - \frac{a}{W})(2.15 - 3.93\frac{a}{W} + 2.7\left(\frac{a}{W}\right)^2)]}{2(1 + 2\frac{a}{W})(1 - \frac{a}{W})^{3/2}} \quad (5)$$

To measure the strength, the ceramics were machined into beams with a length of 30 - 20 mm, thickness of W = 3 - 5 mm and width of B = 4 - 6 mm (B > W), the bottom surfaces were polished into ~1 μm and the edges were chamfered. Three-point fracture bending tests assisted by *in situ* optical microscopy imaging were carried out to measure the sample deflection. The optical setup consists of a 31-megapixel camera with a high speed 10 Gbit/s interface (Ory 10GigE, Teledyne FLIR) and a high resolution telecentric lens (resolution ~5 μm/px, VS-LTC3.3-45/FS, VS Technology). The beams were illuminated with a low angle diffused ring light (VL-LRD73100W, VS Technology), and the images were collected at 1.98 Hz. The sample bending displacements 3D drifting[102] are corrected by ImageJ[100]. The strength $\sigma_f$ and strain $\delta$ were calculated following:

$$\sigma_f = \frac{3FL}{2BW^2} \quad (6)$$

$$\varepsilon = \frac{6\delta W}{L^2} \quad (7)$$

The Young's modulus was determined by the slope of the stress-strain curve. The hardness was determined by Vickers hardness (Indentec, ZwickRoell) under 5 kg and holding for 10 s on well-polished samples (~1 μm).

Piezo-, ferro- and dielectric testing

The samples for functional characterizations were machined out from the sintered disks, made into round disks with diameters of 3-5 mm and thickness of 0.4-0.5 mm. Electrodes were deposited onto the upper and lower end faces of the disk-shaped specimens as follows. First, a layer of conductive silver paste (FuelCellMaterials, FCM 233006 Ag-I) was applied to both end faces and preliminarily dried at 120 °C for 20 minutes. Thereafter, an additional coat of silver paste was applied, and the samples were fired at 650 °C for 30 minutes to ensure robust adhesion of the electrodes.

To measure the piezoelectric properties of the ceramics, samples were poled in a silicone oil bath by applying a DC electric field of 7.5 kV/mm at 80 °C for 30 minutes. A ferroelectric testing system with a laser interferometer (TF1000, aixACCT Systems GmbH, Germany) was used to evaluate the field-induced electromechanical (strain-field and polarisation-field) response. The piezoelectric coefficient $d_{33}$ was measured using a $d_{33}$ meter (PM3001, KCF Technologies, Inc., USA). At least 5-10 samples were used for each characterization. The bipolar fatigue testing was done on the TF1000 with a triangular wave form with amplitude of 7.5 kV/mm and a frequency of 1000 Hz and hysteresis tested at 10 Hz, at least 3-4 samples were used for each characterization.

Energy harvester system installation and output testing

For the setup of the cantilever beam-based piezoelectric transducer, the B&M BNT and BNT reference beams were machined into sheets with width of 5 mm, length of 12 - 13 mm and thickness of 0.9 - 1.1 mm. Each ceramic sheet was glued close to the fixed end of an aluminum cantilever beam with length of 100 mm, width of 10 mm and thickness of 1.2 mm. The fixed end was mounted to an electromagnetic shaker capable of providing one-dimensional reciprocating motion under constant voltage (constant force) with variable-frequency output. And the free end of the cantilever beam was connected to an iron block of mass 16.7 g. The electromagnetic shaker was operated using a sinusoidal waveform under constant driving voltage input at variable sinusoidal frequencies of 2 to 26 Hz. The voltage output of the transducer was recorded using an electrometer (B2987A, Keysight, internal electrical impedance > 1 TΩ), operating in high-impedance mode for open-circuit measurements.

## Acknowledgements

The authors would like to thank Prof. Kyle Webber for fruitful discussion on BNT and lead-free piezoceramics mechanical properties.

## Supplementary methods

Zeta potential measurements

The test was done by ZETASIZER NANO from Malvern. All the powder samples were diluted in 0.05g/ml, and 3ml liquid of this suspension system were transferred into the detection vessel with denoised water (Folded Capillary Zeta Cell, DTS1070).

Archimedes density

The density was measured using the Archimedes' principle, with a laboratory density test instrument Sartorius Quintix Analytical Balance and Sartorius YDK03 Density Kit with pure water. The calculation was based on the following formula: $\rho = \frac{m_1}{m_1 - m_2} * \rho_{pure\ water}$. Where is the $\boldsymbol{m_1}$ mass of the sample in air, $\boldsymbol{m_2}$ is the mass of the sample in deionized pure water, and $\boldsymbol{\rho}$ is the density for sample.

Rietveld analysis for lattice parameters measurements

XRD measurements were performed using an MPD Panalytical XRD diffractometer with a step size of 0.015°, and a scan time of 60 s per degree. The collected data were refined using the Le Bail and Rietveld methods implemented in the GSAS software. Two identical BNT phases were needed to refine the BNT diffractogram: a stress-free BNT phase representing the undistorted structure (named Matrix area), and a stress-affected BNT phase exhibiting pronounced lattice distortion (named Distorted area). All the refinement covariance were wR < 9.5%.

Impedance spectroscopy

The samples are machined into small disks with thickness of 0.4 - 0.5 mm and diameter of 3 - 5 mm, surfaces are grinded and polished into 6 μm. Electrodes were deposited onto the upper and lower end faces of the disk-shaped specimens as follows: first, a layer of conductive silver paste (FuelCellMaterials, FCM 233006 Ag-I) was applied to both end faces and preliminarily dried at 120 °C for 20 minutes; Thereafter, an additional coat of silver paste was applied, and the samples were fired at 650 °C for 30 minutes to ensure robust adhesion of the electrodes. The frequency and temperature evolution permittivity performance were measured by 1296A Dielectric Interface and analysis software SMaRT (from Solartron analytical) from 25 °C to 500 °C under $10^2$ - $10^6$ Hz. The $\varepsilon_{33}^T$ and Loss is calculated based on the formula:

$$\varepsilon_{33}^T = \frac{1}{\varepsilon_0} \times \frac{Cd}{A} \quad (1)$$

$$C = \frac{-1}{2|Z''|\pi f} \quad (2)$$

$$Loss = tan\delta \quad (3)$$

$\varepsilon_0 = 8.85 \times 10^{-12}\ F/m$ is the vacuum permittivity, C is the capacitance calculated based on EIS, d is the sample thickness, A is the sample area, $|Z''|$ is the magnitude of the imaginary part of the impedance in EIS, f is the measuring frequency, $\delta$ is the loss angle.

# Supplementary Figures and Tables

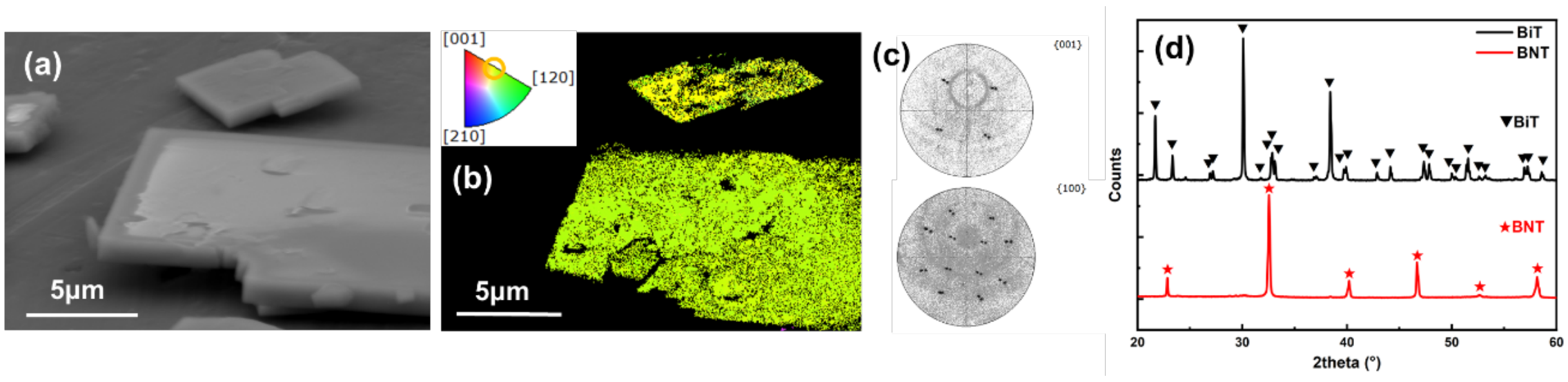

**Figure S1. SEM and EBSD for BNT platelets and single crystal orientation.** (a) BNT platelets SEM morphology. (b) EBSD image pattern for BNT single crystal platelets in (a) with lattice orientation. (c) Pole figure based on EBSD data in (b) at {001} top view and {100} side view observation direction, respectively. (d) X-ray diffraction patterns for BiT templates and BNT platelets during molten salt synthesis and topochemical conversion.

**Table S1.** Zeta potential for BNT platelets evolution for 1vol%$SiO_2$-coated and ~0.1wt% $Fe_3O_4$.

| Powder | Zeta potential (mV) |
|---|---|
| Pure BNT | -45.0 |
| Functionalized BNT (BNT + nano-$Fe_3O_4$ ) | -31.6 |
| Coated & Functionalized BNT (BNT + nano-Fe3O4 + nano-$SiO_2$) | -40.2 |

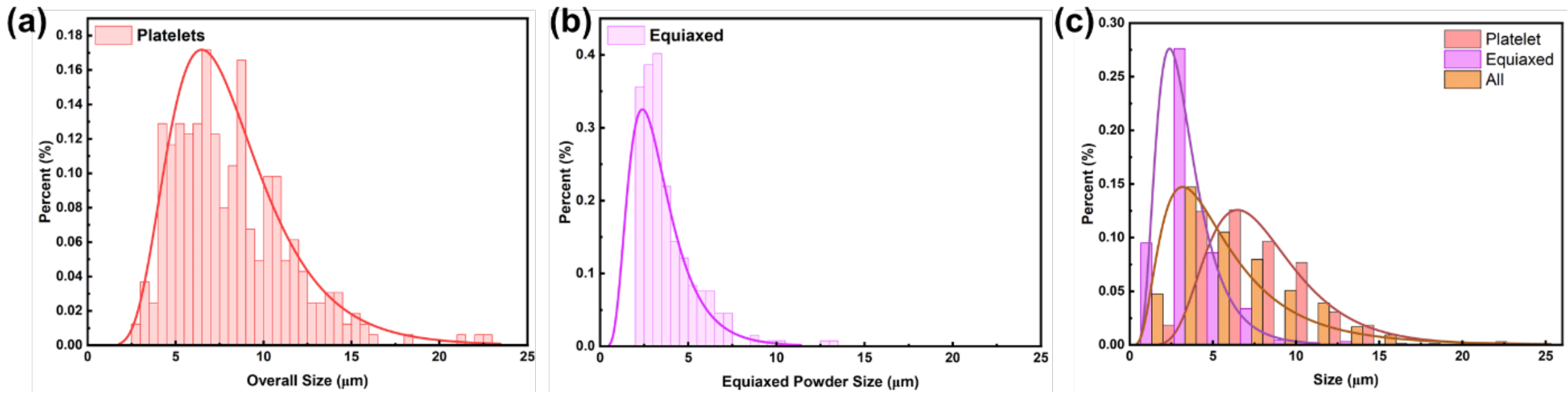

**Figure S2. BNT powder size and shape distribution with platelets diameter and equiaxed powder diameter.** (a) Platelets powder distribution. (b) Equiaxed powder distribution. (c) Overall distribution. The data is fitted with a lognormal function.

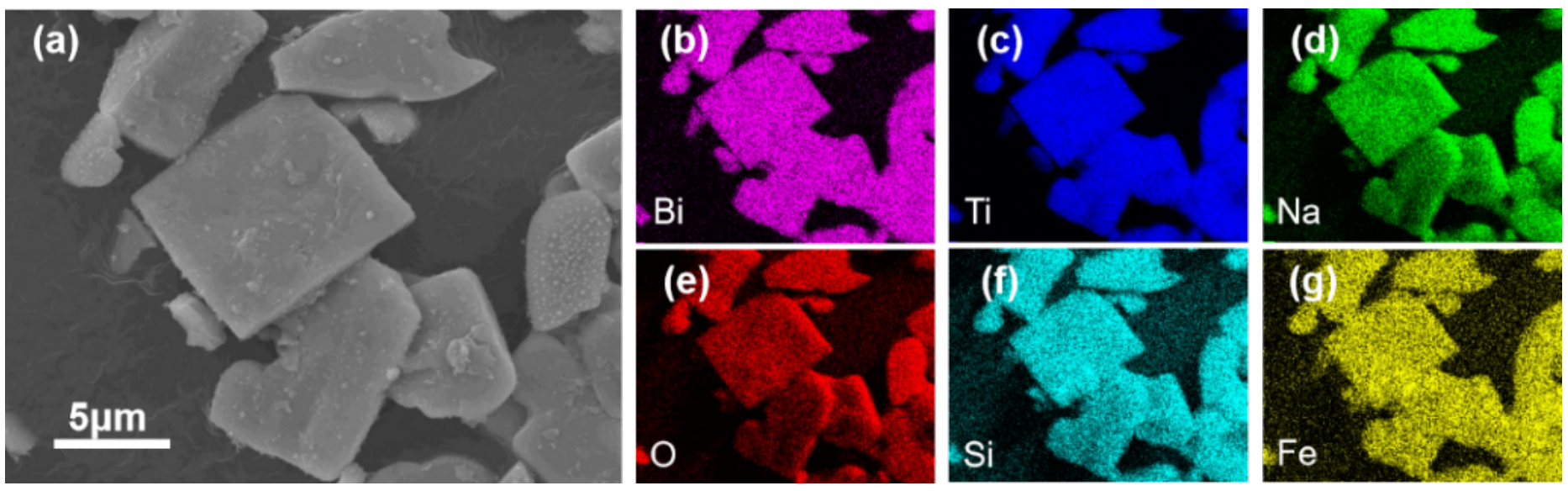

**Figure S3. SEM and EDX for BNT platelets coated by super-magnetic $Fe_3O_4$ and nano-silica.** (a) BNT platelets SEM morphology. EDX elements distribution morphology for (a) Bi, (b) Ti, (c) Na, (d) Ti, (e) O, (f) Si and (g) Fe.

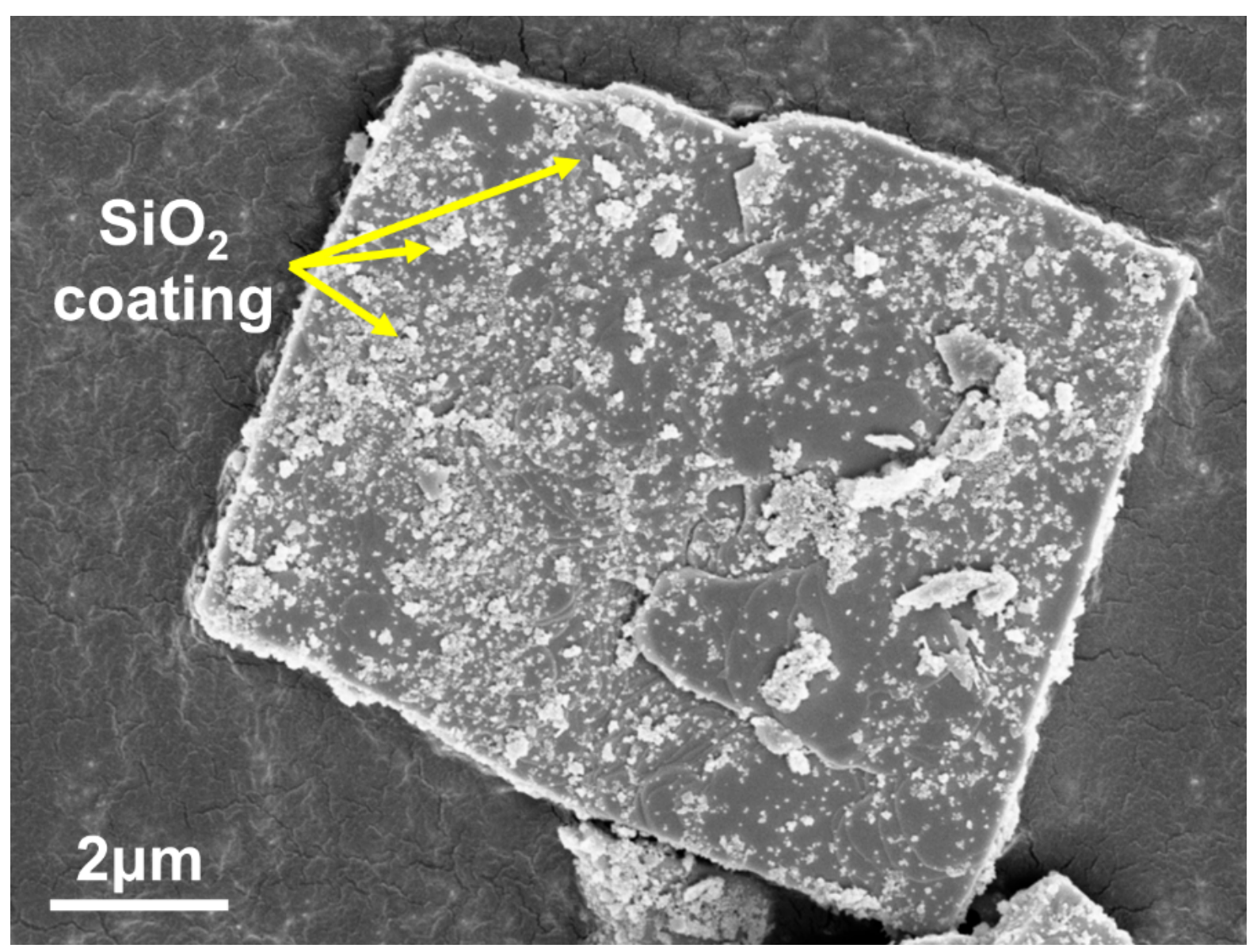

**Figure S4.** SEM image for nano-silica on BNT platelets with 37.5μL/g and 5vol% $SiO_2$.

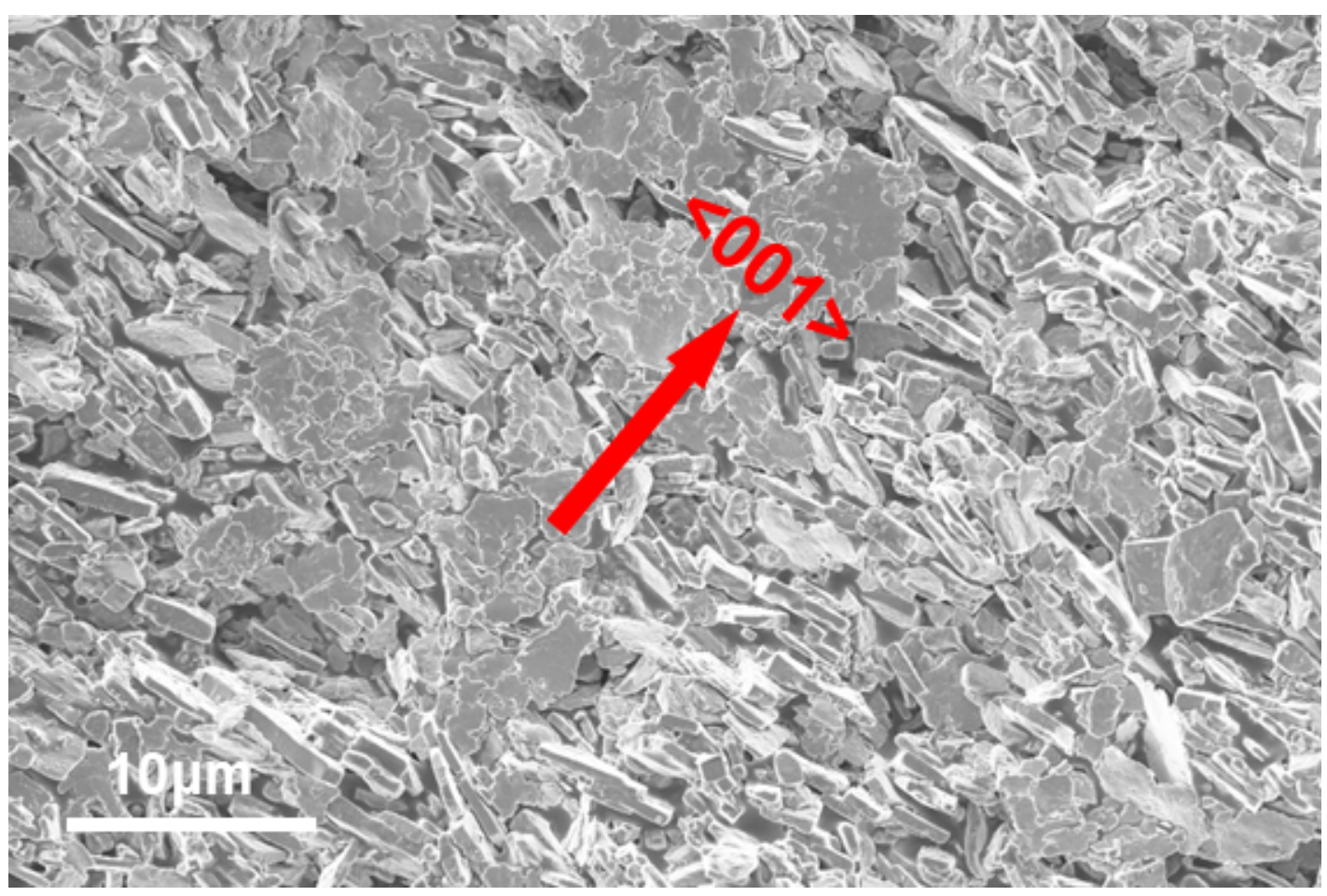

**Figure S5. SEM image for green body cross section with texturing B&M BNT 1vol% $SiO_2$.** The texturing direction <001>pc along platelets thickness direction is marked in the figure. Part of the surface artifacts are lamellar pull-out and contaminants generated during slicing.

**Table S2.** BNT ceramic samples' density.

| Sample | Relative Density ($SiO_2$+BNT) |
|---|---|
| BNT reference normal | 99.6% |
| B&M BNT 0 vol% $SiO_2$ | 99.0% |
| B&M BNT 0.15 vol% $SiO_2$ | 97.6% |
| B&M BNT 0.5 vol% $SiO_2$ | 99.3% |
| B&M BNT 1 vol% $SiO_2$ | 98.2% |
| Equiaxed BNT 1 vol%$SiO_2$ | 97.2% |
| B&M BNT 3 vol% $SiO_2$ | 98.9% |
| B&M BNT 5 vol% $SiO_2$ | 98.5% |

**Table S3.** All textured BNT sample Lotgering factor.

| Sample | Lotgering factor $(00l)_{pc}$ |
| --- | --- |
| B&M BNT 0 vol% $SiO_2$ (green body before SPS) | 95.7% |
| B&M BNT 0 vol% $SiO_2$ | 82.3% |
| B&M BNT 1 vol% $SiO_2$ | 85.7% |
| B&M BNT 3 vol% $SiO_2$ | 89.6% |
| B&M BNT 5 vol% $SiO_2$ | 84.2% |

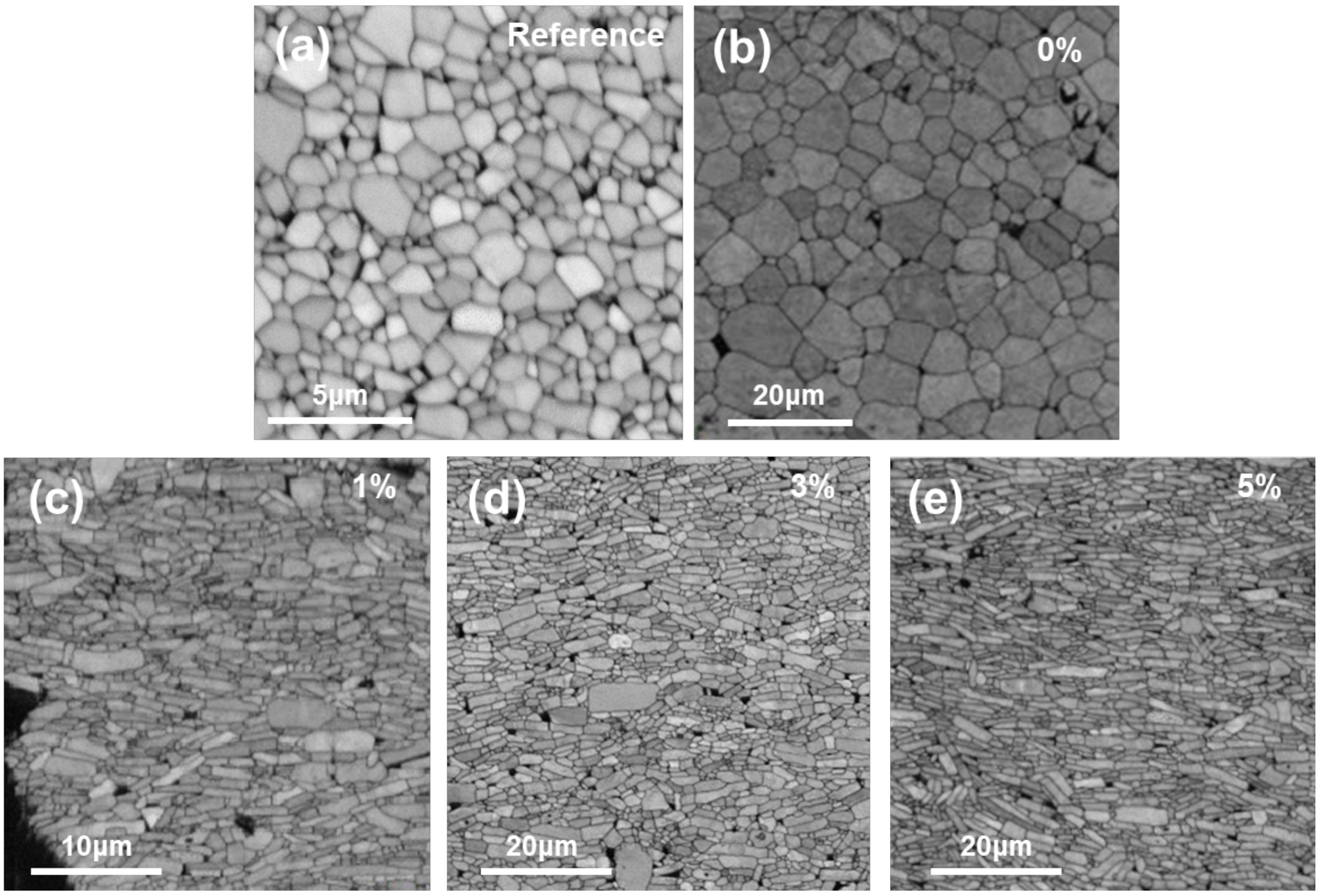

**Figure S6. The microstructure of the BNT-$SiO_2$ composite systems.** EBSD image quality (IQ) pattern for (a) reference BNT, (b) BNT-0%$SiO_2$, (c) BNT-1%$SiO_2$, (d) BNT-3%$SiO_2$, and (e) BNT-5%$SiO_2$, respectively.

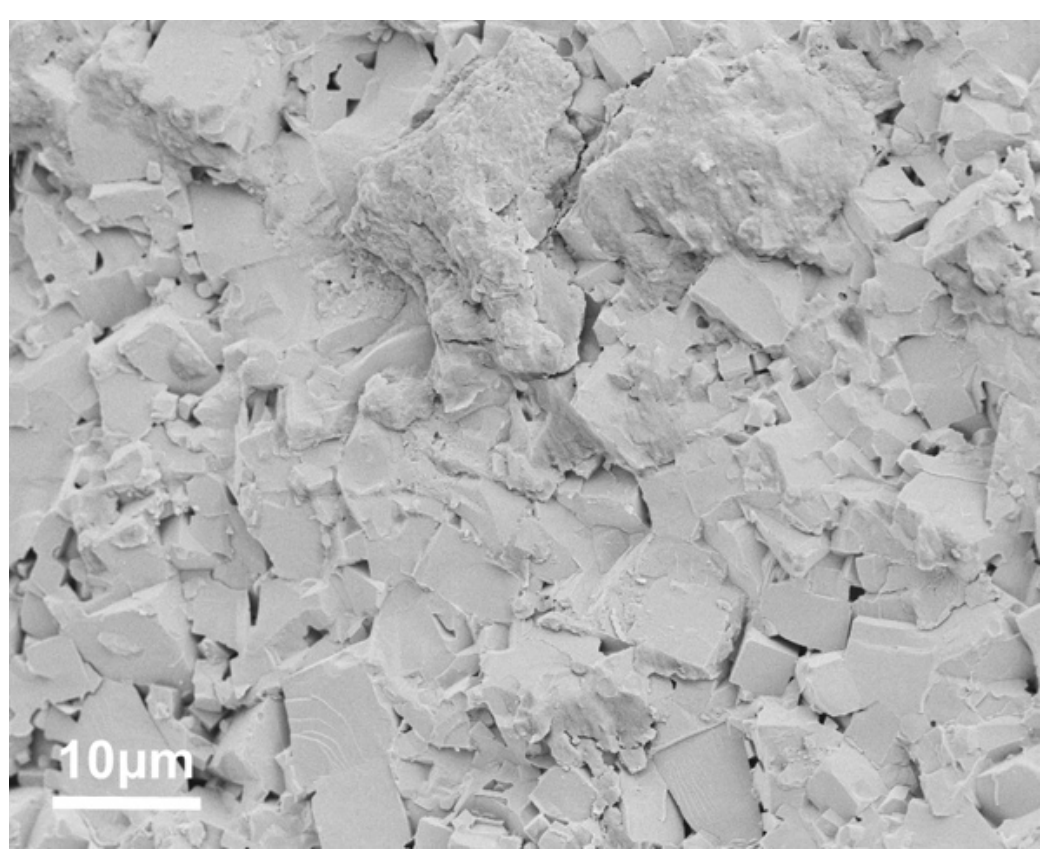

**Figure S7.** Fracture morphology at sample middle part of three-point bend strength test for untextured BNT-1 vol% $SiO_2$ sintered by SPS under 1030 °C and 60MPa.

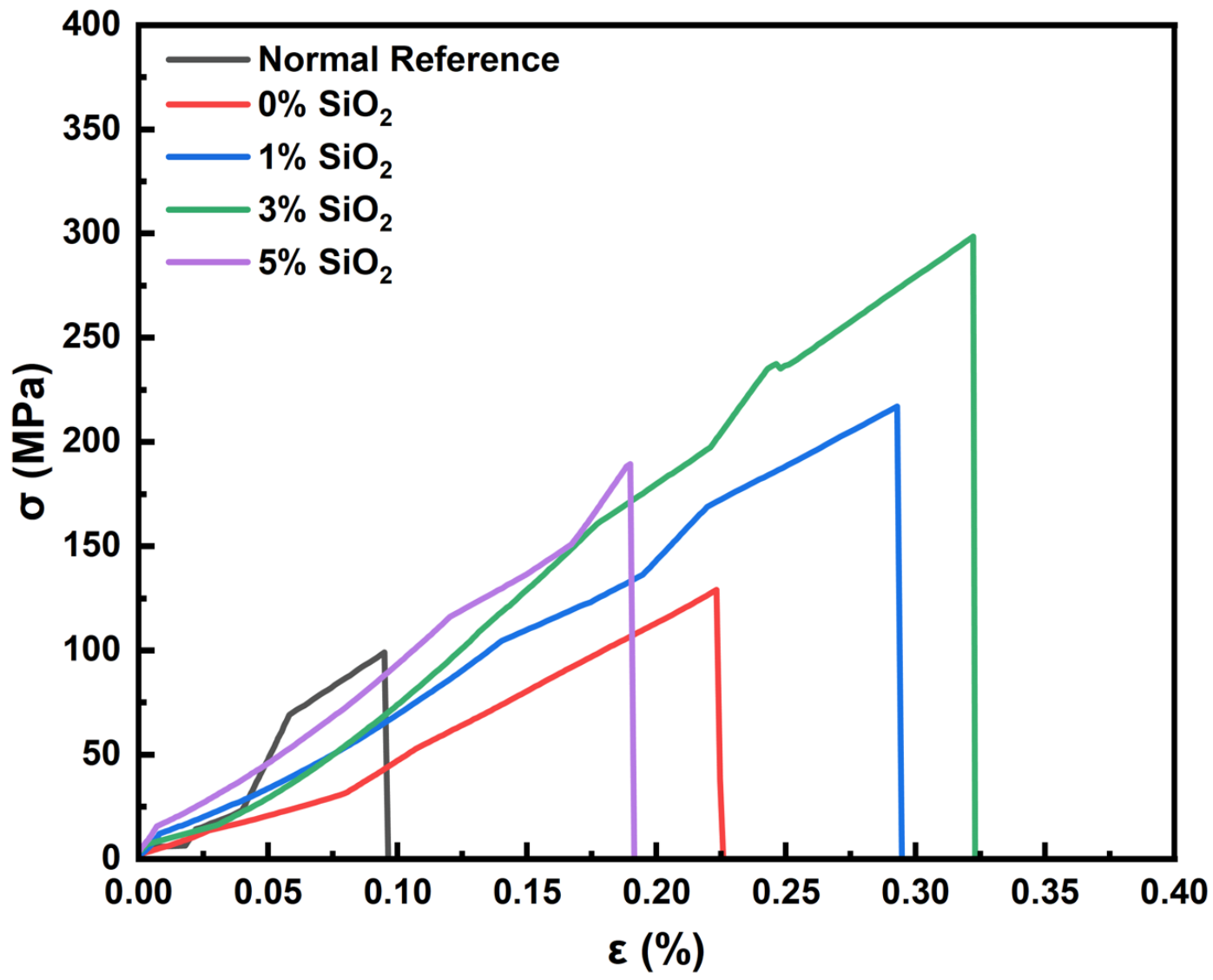

**Figure S8.** Stress-strain curve with optically corrected displacement of the composition with 1-5 vol% of silica compared with the reference BNT.

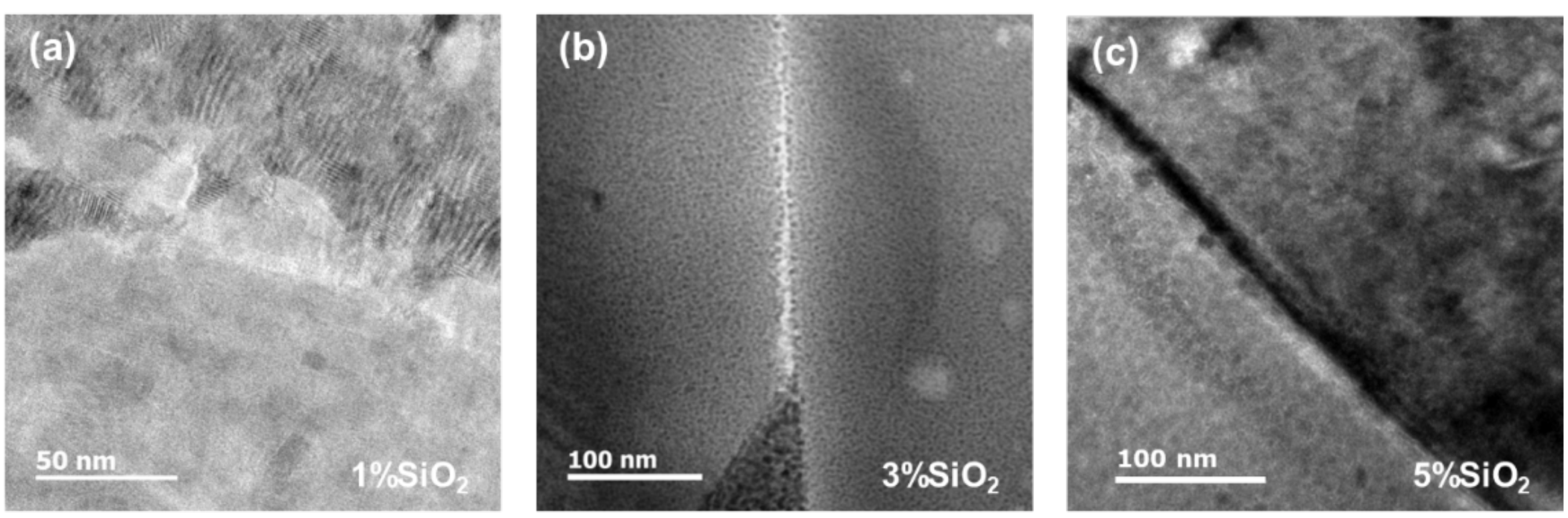

**Figure S9.** TEM figures for parallel platelets grain boundary with $SiO_2$ filled inside for B&M BNT with 1 vol% (a), 3 vol% (b) and 5 vol% (c) $SiO_2$.

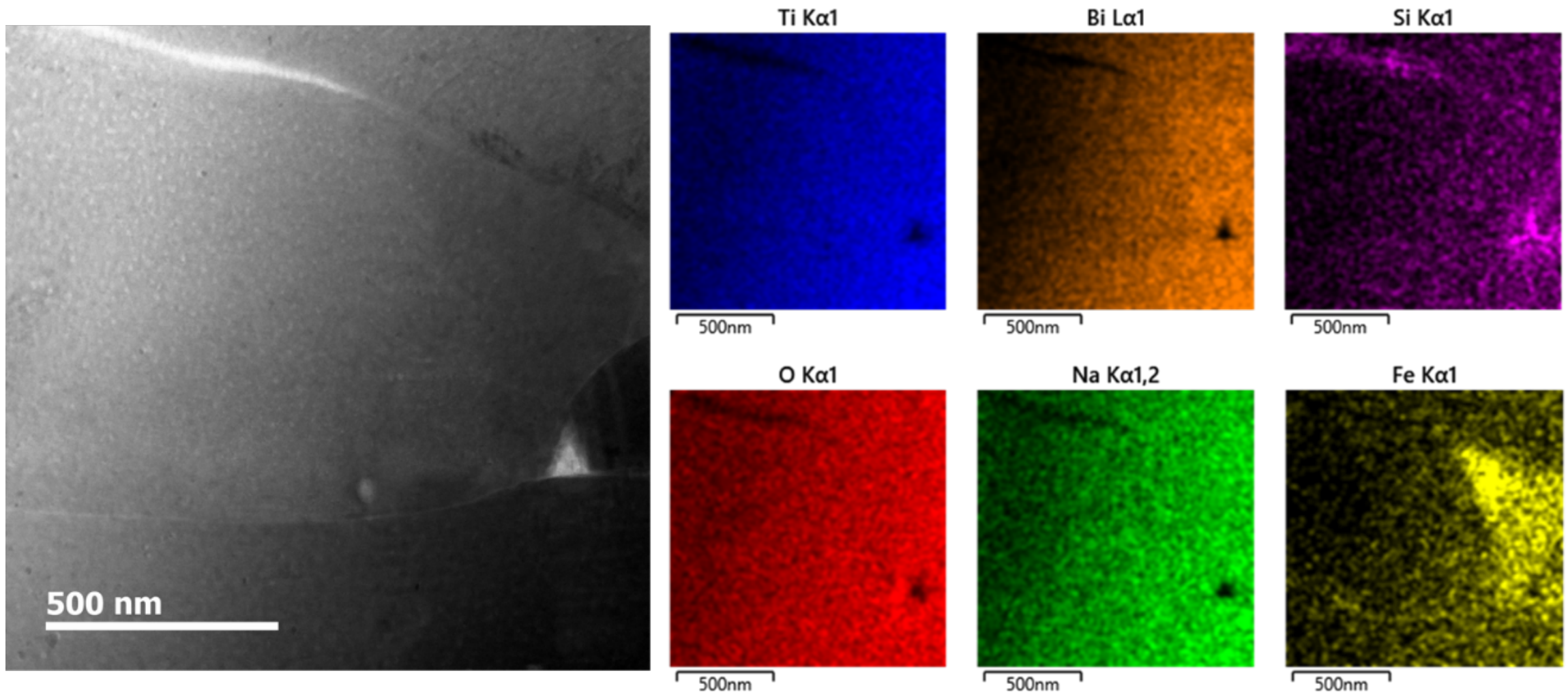

**Figure S10.** TEM and related EDS figures for parallel platelets grain boundary with $SiO_2$ filled inside for B&M BNT with 1 vol% $SiO_2$ with Ti, Bi, Si, O, Na, Fe respectively.

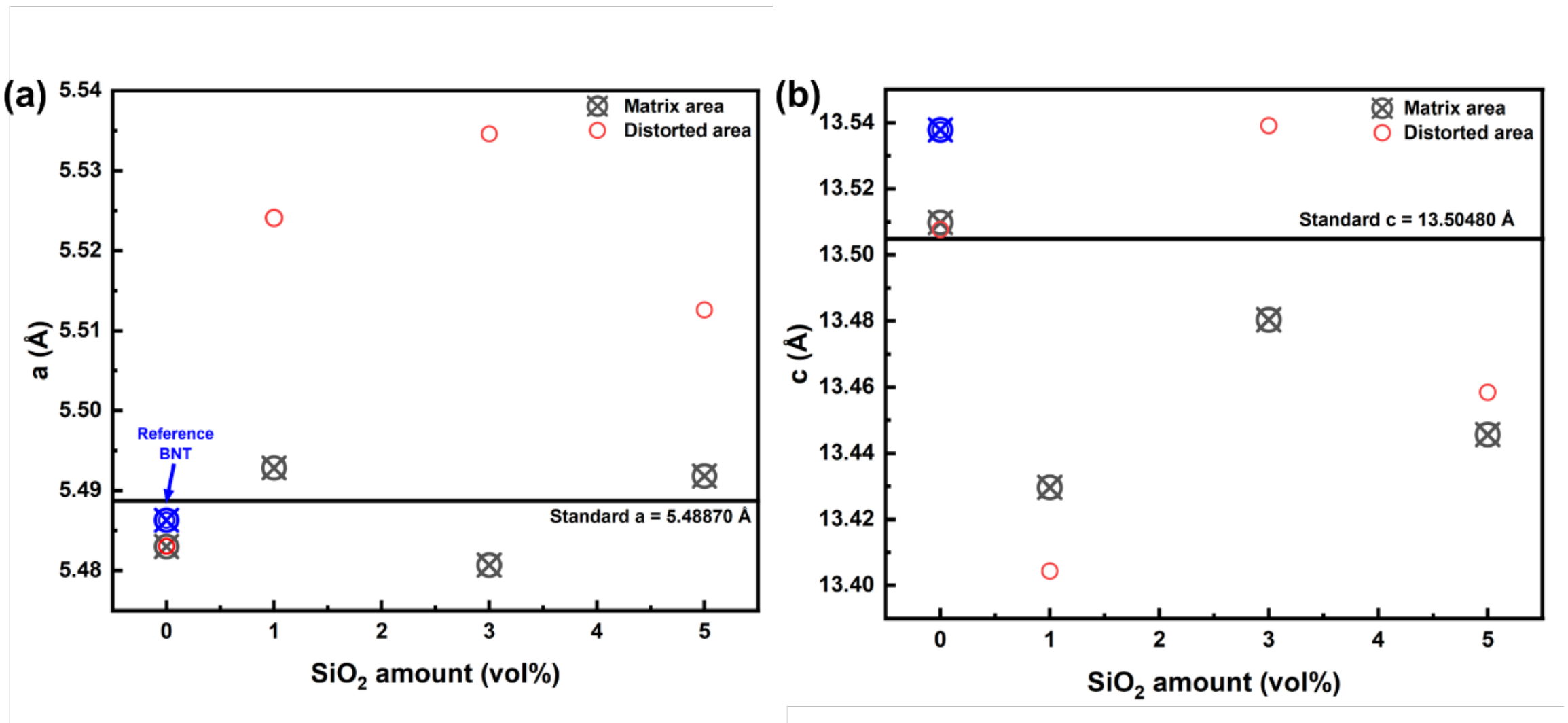

**Figure S11. XRD Rietveld refinement data for reference BNT and B&M BNT samples**. Two identical BNT phases were needed to refine the diffractogram: one corresponding to the BNT with less or no stress; another corresponding to the area under larger residual stress resulting the peak shifting and broadening. The lattice parameters evolving for a (a) and c (b).

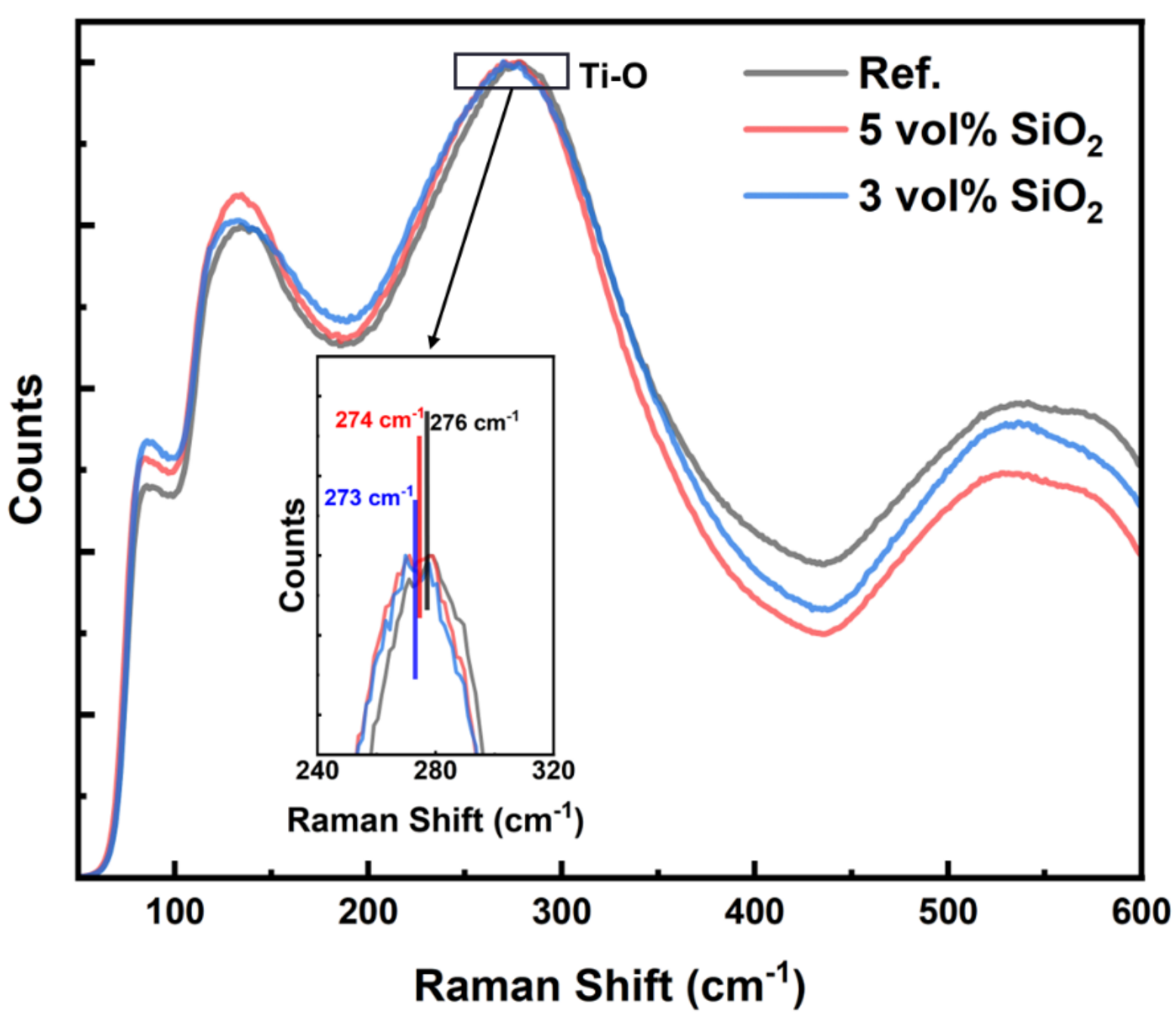

Figure S12. The Raman spectrogram of BNT reference sample and B&M BNT 3 and 5 vol% $SiO_2$, including the peak position of the A1 symmetry related to the $TiO_6$ octahedra.

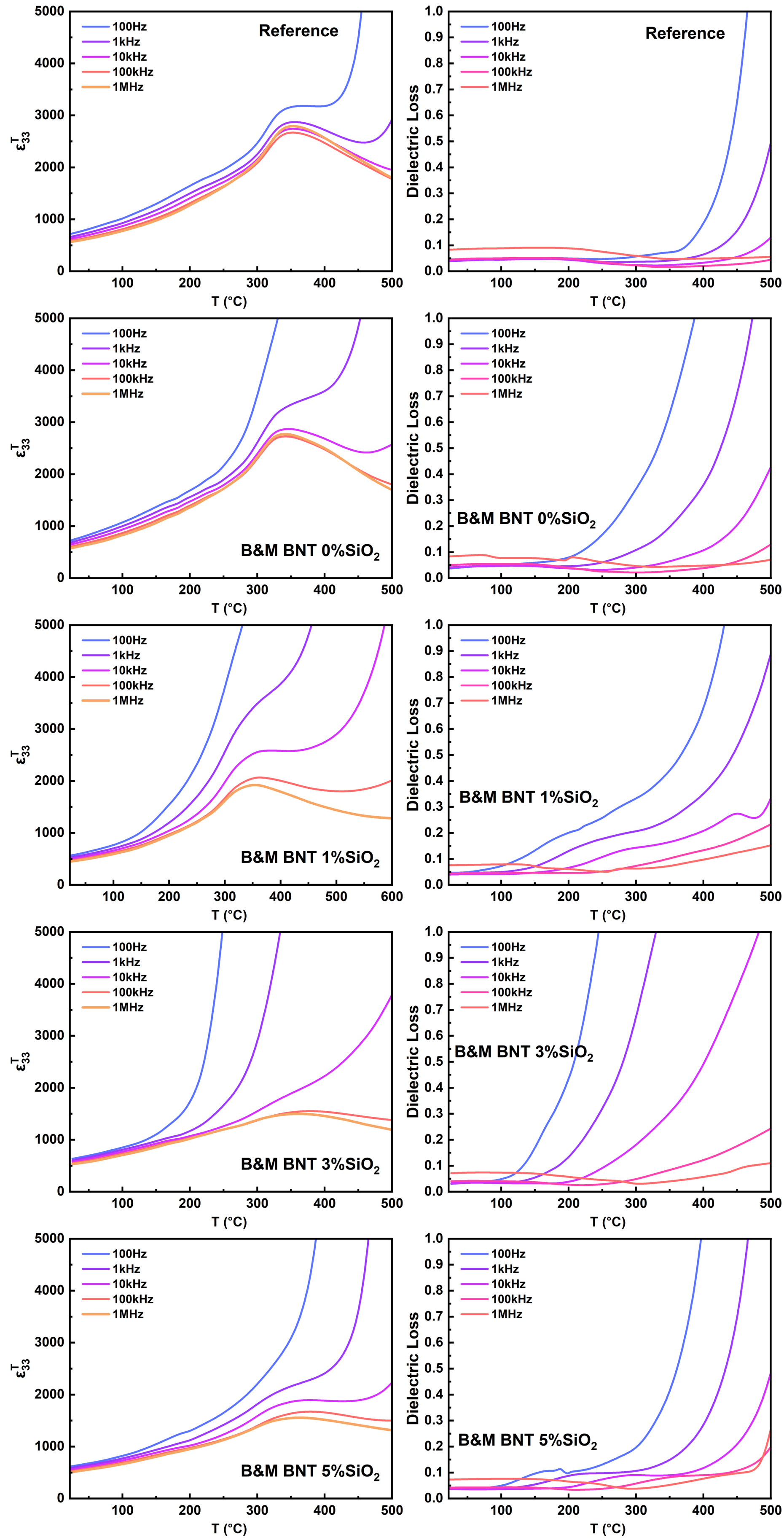

**Figure S13. Collection for Permittivity-Temperature and Loss-Temperature curves based on electrochemical impedance spectroscopy (EIS).** The measurement directions of these relative permittivity $\varepsilon_{33}^{T}$ and loss are the thickness directions of the disks corresponds to the <001>$_{pc}$-direction, $\varepsilon_{33}^{T}$ for reference BNT (a), B&M BNT 0 vol% $SiO_2$ (c), B&M BNT 1 vol% $SiO_2$ (e), B&M BNT 3 vol% $SiO_2$ (g) and B&M BNT 5 vol% $SiO_2$ (i); loss for reference BNT (b), B&M BNT 0 vol% $SiO_2$ (d), B&M BNT 1 vol% $SiO_2$ (f), B&M BNT 3 vol% $SiO_2$ (h) and B&M BNT 5 vol% $SiO_2$ (j).

| | $Bi_4Ti_3O_{12}$ | | $0.94Bi_{0.5}Na_{0.5}TiO_3$-$0.06BaTiO_3$ | | $BaTiO_3$ | |
|---|---|---|---|---|---|---|
| | Random equiaxed system ref. | B&M with 1vol% $SiO_2$ mortar | Random equiaxed system ref. | B&M structure without mortar phase | Random equiaxed system ref. | B&M structure without mortar phase |
| $K_{IC}$ ($MPa*m^{0.5}$) | 0.8 ± 0.2 | 2 ± 0.2 | 0.8 ± 0.05 | 1.0 ± 0.05 | 0.6 - 1.2 | 1.7 ± 0.4 |
| Strength (MPa) | 50 ± 20 | 90 ± 20 | 64 ± 8 | 100 ± 10 | | |
| $d_{33}$ (pC/N) | 6 | 14 | 96 | 104 | | |
| | pole at 8 kV/mm at 120 °C | | pole at 5 kV/mm at room temperature | | | |

**Table S4. Other Nacre-like structured piezoceramics examples with better mechanical performance and better piezoelectric constants in our work to prove the broad viability of this design strategy.** The matrix brick phases include $Bi_4Ti_3O_{12}$, $0.94Bi_{0.5}Na_{0.5}TiO_3$-$0.06BaTiO_3$ and $BaTiO_3$, the mortar phase include $SiO_2$, and normal grain boundary.